\documentclass[12pt]{article}
\usepackage{epsfig,amssymb,amsmath,psfrag,subfigure,rotate,color,wasysym}

\allowdisplaybreaks
\newcommand{\insertfig}[2]{\mbox{\epsfxsize=#1cm \epsfbox{#2.eps}}}

\def\XXint#1#2#3{{\setbox0=\hbox{$#1{#2#3}{\int}$ }
\vcenter{\hbox{$#2#3$ }}\kern-.6\wd0}}

\def \be  {\begin{equation}}
\def \ee  {\end{equation}}
\def \ba  {\begin{eqnarray}}
\def \ea  {\end{eqnarray}}
\def \baa {\begin{eqnarray*}}
\def \eaa {\end{eqnarray*}}
\def \lab #1 {\label{#1}}

\newcommand\re[1]{(\ref{#1})}
\def\d{\hbox{{d}\kern-.20em\hbox{l}}}

\def \matrix #1 {\left(\begin{array}{cc} #1 \end{array}\right)}

\def \tr {\mathop{\rm tr}\nolimits}

\def \e  {\mathop{\rm e}\nolimits}

\newcommand \vev [1] {\langle{#1}\rangle}
\newcommand \VEV [1] {\left\langle{#1}\right\rangle}

\newcommand{\ft}[2]{{\textstyle\frac{#1}{#2}}}
\begin{document}

\begin{titlepage}

\thispagestyle{empty}

\vspace*{2cm}

\centerline{\large \bf An off-shell Wilson loop}
\vspace*{1cm}

\centerline{\sc A.V. Belitsky$^a$, V.A. Smirnov$^{b,c}$}

\vspace{15mm}

\centerline{\it $^a$Department of Physics, Arizona State University}
\centerline{\it Tempe, AZ 85287-1504, USA}

\vspace{5mm}

\centerline{\it $^b$Skobeltsyn Institute of Nuclear Physics, Moscow State University}
\centerline{\it 119992 Moscow, Russia}

\vspace{5mm}

\centerline{\it $^c$Moscow Center for Fundamental and Applied Mathematics}
\centerline{\it 119992 Moscow, Russia}

\vspace{20mm}

\centerline{\bf Abstract}

\vspace{5mm}

It is well-known that on-shell maximally helicity-violating gluon scattering amplitudes in maximally supersymmetric Yang-Mills theory are dual to a bosonic Wilson 
loop on a null-polygonal contour. The light-like nature of the intervals is a reflection of the mass-shell condition for massless gluons involved in scattering. Presently,
we introduce a Wilson loop prototype on a piece-wise curvilinear contour that can be interpreted in the T-dual language to correspond to nonvanishing gluon
off-shellness. We analyze it first for four sites at one loop and demonstrate that it coincides with the four-gluon amplitude on the Coulomb branch. Encouraged by this fact,
we move on to the two-loop order. To simplify our considerations, we only focus on the Sudakov asymptotics of the Wilson loop, when the off-shellness goes to zero. The latter 
serves as a regulator of short-distance divergences around the perimeter of the loop, i.e., divergences when gluons are integrated over a small vicinity of the Wilson loop cusps. 
It does not however regulate conventional ultraviolet divergences of interior closed loops. This unavoidably introduces a renormalization scale dependence and thus scheme 
dependence into the problem. With a choice of the scale setting and a finite renormalization, we observe exponentiation of the double logarithmic scaling of the Wilson loop 
with the accompanying exponent being given by the so-called hexagon anomalous dimension, which recently made its debut in the origin limit of six-leg gluon amplitudes. This 
is contrary to the expectation for the octagon anomalous dimension to rather emerge from our analysis suggesting that the current object encodes physics different from the 
Coulomb branch scattering amplitudes.

\end{titlepage}

\setcounter{footnote} 0

{\small \tableofcontents}

\renewcommand{\thefootnote}{\arabic{footnote}}

\section{Introduction}

On-shell scattering amplitudes occupy the central stage of perturbative QCD studies for almost half a century \cite{Bern:2007dw} since its inception. 
Their infrared behavior is driven by copious gluon emissions and cause their divergence. The latter can be tamed by introducing a small momentum 
cutoff. Of course, physical amplitudes are free from this artificial scale since the gluons hadronize 
into jets of particles observed in detectors replacing the cutoff with a small physical mass parameter via the well-known Bloch-Nordsieck and Kinoshita-Lee-Nauenberg 
theorems. The origin of the aforementioned cutoff, can either come from an artificial gluon mass introduced in internal gluon propagators or from taking external gluons 
off their mass shell and giving them a small virtuality. We will designate both with the same symbol $m$ for the qualitative discussion that follows. The choice for one 
or another is made depending on circumstances where corresponding amplitudes emerge. Both regularizations appear in various factorization schemes between 
ultraviolet and infrared physics as applied to strongly interacting physical processes, the so-called collinear and transverse momentum schemes are the two well-known 
examples of the use of on- and off-shell amplitudes, respectively.

The soft gluon radiation is encoded in the so-called Sudakov form factor of a simpler problem of a two-particle decay of an off-shell gauge boson with momentum $Q$. 
In Abelian gauge theories, the problem of its infrared behavior was solved exactly a long time ago by resumming double logarithmic dependence of the amplitude at each 
order of perturbation series. For on-shell decay products, it is\footnote{We cite the papers on the Sudakov form factor in anti-chronological order since Sudakov studied,
in fact, the off-shell regime.} \cite{Jackiw:1968zz}
\begin{align}
F \sim \exp\left( - \frac{e^2}{8 \pi^2} \log^2 (Q^2/m^2) \right)
\, .
\end{align}
In their off-shell regime, the exponent was found to be twice larger \cite{Sudakov:1954sw}. Further studies unravelled the source for this `discrepancy'. It was 
attributed to the non-vanishing contribution from ultra-soft integration in addition to the soft regime, present in both cases, as explained in Refs.\ 
\cite{Fishbane:1971jz,Mueller:1981sg,Korchemsky:1988hd}.

The above Sudakov form factors are intrinsic building blocks of on-shell non-Abelian scattering amplitudes. Perturbative structure of the latter in QCD is a topic of practical 
phenomenological importance with application to collider physics. The coupling dependence of the infrared asymptotics in this case does not stop at one loop and is driven 
by the so-called cusp anomalous dimensions \cite{Polyakov:1980ca,Korchemsky:1987wg},
\begin{align}
\frac{e^2}{8 \pi^2} \to \frac{1}{2} \Gamma_{\rm cusp} (g_{\scriptscriptstyle\rm YM})
\, ,
\end{align}
depending on the Yang-Mills (YM) coupling $g_{\scriptscriptstyle\rm YM}$. Recent advances in unitarity-based techniques allowed one to reach very high accuracy for 
parton-level amplitudes \cite{Bern:2007dw}. 

The problem of analytical studies of multigluon amplitudes simplify in QCD's supersymmetric cousins, on the one hand, and provide valuable lessons for real physics, one the 
other. The past fifteen years had witnessed an explosion of activity of their exploration within the framework of a distant cousin of QCD, the planar maximally supersymmetric  
YM theory. This is where the gauge/string correspondence \cite{Maldacena:1997re} endows one with the power to probe all values of 't Hooft coupling 
$g^2 = g_{\scriptscriptstyle\rm YM}^2 N_c/(4\pi)^2$ ranging from weak to strong. A major breakthrough that led to this development was an observation of the duality between 
maximally helicity violating on-shell gluon scattering amplitudes  $\mathcal{A}^{\rm on-shell}_N$ and bosonic Wilson loops in the fundamental representation of  SU$(N_c)$ 
gauge group, \cite{Alday:2007hr,Drummond:2007aua,Brandhuber:2007yx}
\begin{align}
\label{MHV-dual}
\mathcal{A}^{\rm on-shell}_N \sim \frac{1}{N_c} \tr P \exp \left( \oint_{C_N} dx \cdot A \right)
\, ,
\end{align}
on a $N$-polygonal contour $C_N = [x_1, x_2] \cup [x_2, x_3] \cup \dots \cup [x_N, x_1]$ with each segment being given by the light-like gluon momentum $p_j$ 
to echo the gluon mass-shell condition $p_j^2 = 0$.  The T-duality relation $p_j = x_j - x_{j+1} \equiv x_{j j +1}$  automatically solves the energy-momentum conservation condition 
$\sum_j^N p_j = 0$. This duality was confirmed by two-loop analyses for $N=4,5$ point Wilson loops \cite{Drummond:2007cf,DruHenKorSok07}  by confronting these predictions with 
available results for gluon scattering amplitudes \cite{ABDK03,BDS05}. Its most stringent test comes though from the comparison of the two-loop result for $N=6$
Wilson loop \cite{DruHenKorSok08,AnaBraHesKhoSpeTra09,DelDuhSmi10,GonSprVerVol10} with the six-gluon MHV scattering amplitude computed to 
the same order in 't Hooft coupling of the planar theory \cite{BerDixKosRoiSprVerVol08,Cachazo:2008hp}. 

Further, the Wilson loop reformulation of gluon amplitudes  paved a way for the use of heavy machinery of integrable models to solve dynamics of the two-dimensional 
world-sheet stretched on its contour \cite{Basso:2013vsa}. It relies on a dual description of amplitudes in terms of excitations propagating on a color flux-tube sourced by 
the Wilson contour. The vacuum represented by the flux is in fact SL(2) invariant to lowest order in 't Hooft coupling \cite{Gaiotto:2010fk,Basso:2010in} with its energy
density determined by the cusp anomalous dimension. The framework was dubbed the Pentagon Operator Expansion and was used to predict multiloop and strong
coupling amplitudes \cite{Basso:2013aha,Basso:2014koa,Belitsky:2014sla,Belitsky:2015efa,Belitsky:2016vyq,Basso:2014nra,Belitsky:2014lta,Basso:2014hfa,Basso:2015rta}.

Off-shell scattering received less attention mainly because naively taking external gluons off their mass shell renders their amplitudes gauge dependent. Nevertheless, these 
play a crucial role in QCD studies. Of particular importance to phenomenological applications in hadron-hadron collisions is the so-called high-energy factorization and 
Berends-Giele recursion relations for matrix elements of currents with different number of gluon legs \cite{Berends:1987me}. In the former application, contrary to the more 
conventional collinear factorization, amputated legs of Green's functions after the LSZ reduction are not set to their mass shell\footnote{See also Ref.\ \cite{Ermolaev:2005gu}, 
where the leading double logarithmic asymptotics of $2 \to 2$ off-shell scattering was related to their on-shell limit by studying infrared evolution equations.}.  Gauge dependence 
is an obvious problem but there is a number of possible ways out of this predicament. One consists in a modification of the Feynman diagram technique to automatically include  extra 
contributions that reconstitutes their gauge invariance \cite{Lipatov:1995pn,Kotko:2014aba}. Another is to perform all calculations in a background gauge where the gauge in the 
internal lines can be disentangled from the gauge of the trees and therefore the loop effects can be computed in a gauge invariant fashion as an effective action 
\cite{Gates:1983nr,Boels:2010mj}. Finally, the off-shellness can be introduced legitimately in $\mathcal{N}=4$ SYM by considering its spontaneously broken phase by giving 
non-vanishing expectation values to scalar fields and making all external gluon legs massive \cite{Alday:2009zm}. This regime was dubbed the Coulomb branch.

A recent study \cite{Caron-Huot:2021usw} of the four-gluon scattering amplitude on the above Coulomb branch suggested its intriguing connection to a four-point correlation function
of half-BPS operators with infinitely large $R$-changes. It was realized some time ago \cite{Coronado:2018ypq,Coronado:2018cxj}, that for a particular case of the so-called 
simplest four-point correlator, it factorizes into a product of two objects named octagons $\mathbb{O}$. The latter can be cast in a concise representation in terms of a determinant 
of a semi-infinite matrix \cite{Kostov:2019stn,Kostov:2019auq}. This result laid out the foundation for further considerations in Refs.\ \cite{Belitsky:2019fan,Belitsky:2020qrm,Belitsky:2020qir}, 
where the octagon was further recast as a Fredholm determinant of an integral operator acting on a semi-infinite line. Its kernel was identified as a convolution of the well-known 
Bessel kernel with a Fermi-like distribution that depends on the external kinematics and the 't Hooft coupling. In Refs.\ \cite{Belitsky:2019fan,Belitsky:2020qzm}, the null limit of 
the octagon was addressed. It corresponds to the kinematics where any two nearest-neighbor operators approach a light-like interval such that one of the conformal cross ratios, 
i.e., $y$, tends to infinity and all other can be set to zero. One can establish the asymptotic behavior of the octagon as $y\to\infty$ \cite{Coronado:2018cxj}
\begin{align}
\label{NullOctagon}
\mathbb{O} = \exp\left( - \frac{ y^2 }{4} \Gamma_{\rm oct} (g) \right)
\, ,
\end{align}
and exactly fix the accompanying coefficients as functions of the coupling known as the octagon anomalous dimension, \cite{Belitsky:2019fan,Belitsky:2020qzm}
\begin{align}
\label{GammaOct}
\Gamma_{\rm oct} (g) = \frac{2}{\pi^2} \cosh (2 \pi g)
\, .
\end{align}
The authors of Ref.\ \cite{Caron-Huot:2021usw} suggested that the octagon is nothing else as the four-point gluon amplitude on the Coulomb branch. In particular, in the
Sudakov limit, it reads
\begin{align}
\mathcal{A}_4^{\rm off-shell} \sim \mathbb{O}|_{y^2 \to \log^2 \left(m^4/(s t) \right)}
\, .
\end{align}
Further evidence for emergence of the octagon anomalous dimension in the Coulomb branch amplitudes was given Ref.\ \cite{Bork:2022vat} for the five-gluon case. This 
made certain observations made in Ref.\ \cite{Belitsky:2019fan} more transparent. Namely, it was noticed there that one can write two complementary equations which yield 
the octagon anomalous dimension as their solutions: one is based on the BMN vacuum and another one on the GKP one. The origin for the latter was obscure since the 
hexagonalization framework \cite{Basso:2015zoa,Fleury:2016ykk,Eden:2016xvg}, that is used to derive the octagon, is built on the BMN vacuum. The GKP vacuum, on the 
other hand, is related to spinning strings with large angular momentum, which in the dual language corresponds to operators with large Lorentz spin and Wilson lines. 
Let us point out here that a different manifestation of Eq.\ \re{GammaOct} in the origin limit of the on-shell six-gluon amplitude\footnote{Yet another place where the same
flux-tube kernel has made it (unexpected) appearance is the OPE program for $\mathcal{N}=4$ form factors \cite{Sever:2021xga}. We thank Lance Dixon for pointing this 
out to us.} was observed in Ref.\ \cite{Basso:2020xts} building up on the seminal work \cite{Caron-Huot:2019vjl} at seven loop order. In this study, in addition to the already 
discussed soft exponents, yet another one was found to drive the Sudakov-like behavior, which was dubbed the hexagon anomalous dimension $\Gamma_{\rm hex}$ 
\cite{Basso:2020xts}. To make the comparison between them transparent, we quote here their leading two terms in 't Hooft perturbative series 
\begin{align}
\label{ADs}
\Gamma_{\rm \alpha} (g) = 4 g^2 + c_\alpha g^4 + O (g^6)
\, , \qquad c_\alpha = \{ - 16 \zeta_2 , - 8 \zeta_2 , - 4 \zeta_2 \}  
\, ,
\end{align}
for $\alpha = \{ {\rm cusp}, {\rm oct}, {\rm hex} \}$, respectively. 

A natural question which emerges is whether the known duality between on-shell amplitudes and Wilson loops on light-like contours has a generalization to off-shell amplitudes and 
some version of a Wilson loop as well. The flux-tube, however, sourced by color charges traveling around its contour should be different from the one in Eq.\ \re{MHV-dual}, since 
the flux-tube density is different in two cases, i.e., $\Gamma_{\rm cusp}$ vs.\ $\Gamma_{\rm oct}$. Thus, the goal of this paper is to propose a perimeter regularization for the bosonic 
Wilson loop which corresponds to off-shellness of gluons involved in scattering. Getting ahead of ourselves, however, we can report here that what we find is that the Sudakov
limit of our version of the off-shell Wilson loop is driven by the hexagon anomalous dimension rather than the anticipated octagon. We do not have an explanation for this fact.

Our subsequent presentation is organized as follows. In the next section, we introduce an off-shell Wilson loop by means of dimensional reduction of its ten-dimensional progenitor.
As we demonstrate there, the off-shellness requires the Wilson loop contour to reside in more than four dimensions. This naturally introduces a regularization of short-distance cusp
divergences around the perimeter of the loop. The main framework that we employ to calculate loop corrections is the method of Lagrangian insertions. First, we perform a one-loop 
analysis in Sect.\ \ref{SectionOneLoop} of the four-site Wilson loop, demonstrating its exact equivalence to the four-gluon amplitude on the Coulomb branch \cite{Caron-Huot:2021usw}. 
Next, in Sect.\ \ref{OffOnShellSection}, we provide a set of one-loop interpolation formulas between the on- and off-shell regimes by using analytic continuation in the parameter of 
dimensional regularization in the interior of the loop. This agreement fosters further analysis of two-loop effects in Sect.\ \ref{2LoopsSection}. Here, we limit ourselves however only to 
the consideration of the small-virtuality limit which makes calculations sufficiently involved already as the emerging integrands are akin to massive two-loop amplitudes. To this end, we 
rely on the powerful technology of the expansion by regions and Mellin-Barnes integral representations to find asymptotic behavior of all contributing Feynman graphs. We observe 
exponentiation of the double logarithms and determine the functions of the 't Hooft coupling which drives it, the hexagon anomalous dimension. Finally, we conclude. A few appendices 
corroborate some of the findings in the main body of the paper or collect discussion and formulas which were not suited for the main text. 

\section{An off-shell Wilson loop}

Up to now, the number of attempts to devise a version of the Wilson loop that corresponds to off-shell gluons can be counted on the fingers of one hand. Namely,
Ref.\ \cite{Dorn:2008dz} proposed to cut-off the immediate vicinity of the cusps thus breaking gauge invariance of the loop. The leading double logarithmic dependence 
of the four-site loop was shown to be gauge independent however. While Ref.\ \cite{Gorsky:2009dr} proposed to modify the coordinate-space propagators ``by hand" shifting 
Lorentz invariant distances between two points by a mass parameter, i.e., $x^2 \to x^2 + m^2$. While it works at one loop, this off-shell regulator was demonstrated to fail 
anticipated exponentiation property starting from two loops. These two examples exhaust the list of proposals we are aware of.

\subsection{Motivation}

Our goal in this section is to find a holonomy of the gauge connection whose expectation value develops the perturbative expansion which matches the one of an off-shell four-gluon 
amplitude ratio function devised\footnote{Strictly speaking \cite{Caron-Huot:2021usw} did not compute any amplitudes instead they interpreted a 10D light-like limit of a generating 
function of four-point correlators as a scattering amplitude on restricted Coulomb branch of $\mathcal{N} = 4$ SYM with support from multiloop $D$-dimensional integrands of four-leg
amplitudes \cite{Bern:2006ew,Bern:2012uc}. It would definitely be important to calculate off-shell gluon amplitudes directly using, for instance, the background field method.} in Ref.\ 
\cite{Caron-Huot:2021usw}.  Namely, the latter was given to the lowest few orders of the series in 't Hooft coupling $g^2 = g_{\rm YM}^2 N_c/(4 \pi)^2$ by
\begin{align}
\label{OffShellAmplitude}
\mathcal{A}_4^{\rm off-shell}
=
1 
- g^2 \left( z_{13}^2 z_{24}^2 g_{1234} \right)
+ g^4  \left( z_{13}^2 z_{24}^4 h_{13;24} + z_{13}^4 z_{24}^4 h_{24;13} \right)
+ O (g^6)
\, ,
\end{align}
in terms of the standard one- and two-loop ladder integrals\footnote{The one-loop off-shell amplitude in $\mathcal{N} = 4$ SYM was already calculated in terms of the box integral back 
in Ref.\ \cite{Gates:1983nr}, see Eqs.\ (6.5.67) - (6.5.68) there.} rewritten via the dual coordinates as
\begin{align}
\label{CrossIntegrals}
g_{1234} = \frac{1}{\pi^2} \int \frac{d^4 x_0}{x_{01}^2 x_{02}^2 x_{03}^2x_{04}^2}
\, , \qquad
h_{13;24} = \frac{1}{\pi^4} \int \frac{d^4 x_0 d^4 x_{0'} }{(x_{01}^2 x_{02}^2 x_{04}^2) x_{00'}^2 (x_{0'2}^2 x_{0'3}^2 x_{0'4})}
\, .
\end{align} 
In this expression, the external kinematics inherits the full 10D Lorentz symmetry through the invariants\footnote{We use $z$, $x$ and $y$ variables for the 10D, 4D
and 6D coordinates, respectively. Below, we recycle them for the $D$-dimensional, 4 and $(2 \varepsilon)$ spaces, accordingly.} $z_{jj'}^2 \equiv x_{jj'}^2 - y_{jj'}^2$, 
while the loop integrations are performed only on its Minkowski four-dimensional subspace. The nearest-neighbor points in $10$D live on light rays obeying the null conditions 
$x_{jj+1}^2 = y_{jj+1}^2 \equiv m^2_j$ with the extra-dimensional invariants $y_{jj+1}^2$ playing the role of the off-shellness (or external mass). The necessity 
to have massless internal lines forces one to impose another null condition on all $y$-coordinates themselves, i.e., 
\begin{align}
\label{NullYs}
y_j^2 = 0
\, .
\end{align} 
This equation allows one to restore the above $10$D symmetry at the level of the integrand provided the integration variables are localized only in four dimensions  
\cite{Caron-Huot:2021usw}.

\begin{figure}[t]
\begin{center}
\mbox{
\begin{picture}(0,100)(245,0)
\put(0,0){\insertfig{17}{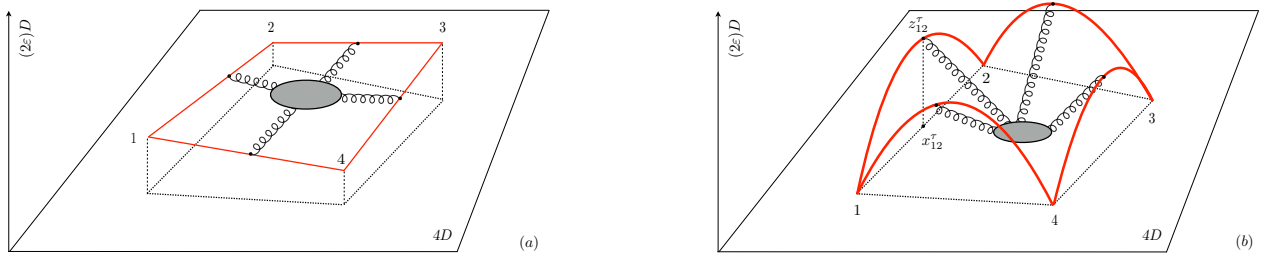}}
\end{picture}
}
\end{center}
\caption{\label{WilsonLoopPic}  Schematic drawing of the $D = 4+ 2\varepsilon$ dimensional contour of the off-shell Wilson loop in (a) and its 
representation in (b) after imposing the light-like conditions \re{NullYs} and \re{LCcondition} and dimensional reduction of the interior to four dimensions.}
\end{figure}

\subsection{Definition}

Having this inspiration in mind, we start with a ten-dimensional bosonic holonomy and dimensionally reduce it down to $D = 4 + 2 \varepsilon$ dimension with 
$\varepsilon > 0$. Equivalently for us, we can view it as a dimensional reconstruction of the $D$-dimensional object from its four-dimensional counterpart\footnote{
If we were to think about this object from the point of view of the four-dimensional Lorentz algebra, the 4D gauge-field components transform as vectors with the 
remaining $2\varepsilon$ being scalars. This would imply that the contour couples to scalars as well, reminiscent of the Maldacena-Wilson loop 
\cite{Maldacena:1998im,Rey:1998ik}. However, we think of our object as truly living just above four dimensions so that it transforms under the so$(D,2)$ 
conformal algebra and all $D$ components of the gauge field are true vectors. This approach is clearly different from the traditional dimensional reduction
 \cite{Siegel:1979wq}.}. Its vacuum expectation value is defined by the path integral
\begin{align}
\vev{W_N} = Z^{-1} \int [dA] \e^{i S} W_N
\, , \qquad
Z = \int [dA] \e^{i S}
\, .
\end{align}
Without loss of generality we will set $Z = 1$ in what follows. The $N$-site Wilson loop defined as
\begin{align}
W_N = \frac{1}{N_c} \tr [w_{[1,N]} \dots w_{[3,2]} w_{[2,1]}]
\end{align}
on a piece-wise contour $[z_1,z_2]\cup [z_2,z_3] \cup \dots \cup [z_N, z_1]$, with links ordered from right-to-left reflecting the path-ordering of fields in the individual 
Wilson segments along the contour
\begin{align}
\label{WLsegment}
w_{[j+1, j]} 
&= P \exp \left( i \int_{z_j}^{z_{j+1}} d z_M  A^M (z) \right)
\\
&=1 + i \int_{z_j}^{z_{j+1}} d z_M  A^M (z) + i^2 \int_{z_j}^{z_{j+1}} d z_M A^M (z) \int_{z_j}^{z} d z'_N A^N (z') + \dots 
\nonumber
\, .
\end{align}
The Boltzmann weight in the path integral is given in terms of the $D$-dimensional action $S = \int d^{D} z_0 \mathcal{L} (z_0)$ of the gauge-fixed Lagrangian\footnote{Only the 
gluon portion is shown with ghosts and other $\mathcal{N} = 4$ fields dropped for the time being. We follow conventions of Ref.\ \cite{Belitsky:2003sh}.}
\begin{align}
\label{TruncLag}
\mathcal{L} = - \frac{1}{g^2_{\scriptscriptstyle\rm YM}} \tr \left( \frac{1}{2} F_{MN}^2 + \frac{1}{\xi} (\partial_M A_M)^2 + \dots \right)
\, .
\end{align}
So far this Wilson loop lives above four dimensions, i.e., its contour as well as its ``interior'' (see the panel $(a)$ of Fig.\ \ref{WilsonLoopPic}). We decompose the $D$-dimensional 
coordinates as $z_j^M = (x_j^\mu, y_j^I)$  in terms of the four-dimensional Minkowski coordinates $x^\mu$ and the auxiliary $(2 \varepsilon)$-dimensional orthogonal 
vector subspace pa\-ra\-metrized by $2 \varepsilon$-component vectors $y^I$. Since vector spaces of non-integer dimensions do not exists, the orthogonal $2 \varepsilon$ 
subspace in fact has to be regarded as infinite dimensional with $I = 1, 2, \dots, \infty$. This is akin to the use of conventional dimensional regularization for calculation of 
divergent field-theoretical integrals where integration rules over fractional dimensions are shown to be consistent and proven upon working in infinite dimensional linear spaces and 
further analytic continuation to other values of $D$, fractional or even complex (see, for instance, the standard text on renormalization \cite{Collins:1984xc}). This `loophole' will 
allow us to accommodate arbitrary complex-valued vectors $y^I$, including those possessing zero norms\footnote{One explicit realization that a reader can imagine for a four-cusp 
loop contour is $y^I_j = \delta_{I,1} + i \delta_{I,j}$.} $(y^I)^2 = 0$ as in Eq.\ \re{NullYs}.

The reason to stay above four dimensions should be quite obvious, we would like to couple propagating gauge fields to the auxiliary coordinates $y^I$, which play a crucial 
role in the introduction of the off-shellness, as we just alluded to above\footnote{Let us note that our setup does not rely on an underlying dual stringy picture, nevertheless
it qualitatively resembles a configuration for Coulomb-branch amplitudes of Ref.\ \cite{Alday:2007hr,Alday:2009zm}, where open strings stretch between D-instantons separated by 
light-like distances at radial distance $r = m$ away from the boundary. We would like to thank Frank Coronado for a discussion of this point.}. Namely, any two adjacent points along 
the polygonal contour in $D$-dimensions (see the panel $(a)$ in Fig.\ \ref{WilsonLoopPic}) form light-like distances $z_{jj+1}^2 = 0$ thus enforcing the constraint 
\begin{align}
\label{LCcondition}
x_{jj+1}^2 = y_{jj+1}^2
\, .
\end{align} 
It is interpreted in light of the T-duality relation for four-dimensional momenta as development of nonvanishing virtualities $y_{jj+1}^2$ for external gluons of scattering 
amplitudes. In what follows, we will assume for simplicity that all of them possess the same value $y_{jj+1}^2 = m^2$.

Parametrizing each $w_{[j+1, j]}$-segment \re{WLsegment} of the  $D$-dimensional contour by the proper time $\tau_j$, the gauge field on each of them resides at the 
position 
\begin{align}
\label{Path}
z_{jj+1}^{\tau_j} \equiv \bar\tau_j z_{j}  + \tau_j z_{j + 1}
\, ,
\end{align}
(here and below $\bar\tau \equiv 1 - \tau$) and is integrated over the interval $0 \leq \tau_j \leq 1$. From the $D$-dimensional perspective\footnote{We would like to thank
Simon Caron-Huot for a discussion of the fate of the conformal symmetry.} the $z_{jj+1}$ interval is light-like \re{LCcondition}, $z_{jj+1}^2 = 0$, so the Wilson loop contour is 
automorphic under the $D$-dimensional conformal inversion, $I[z_j^M] = z_j^M/z_j^2$. Namely, inverting an arbitrary point $z^M = z^M (\tau)$ at some proper time $\tau_j = \tau$ 
on the path \re{Path}, we find
\begin{align}
\label{InvarianceContour}
z^M = \bar\tau z^M_{j}  + \tau z^M_{j + 1} \to I [z^M] = \frac{ \bar\tau z^M_{j}  + \tau z^M_{j + 1}}{\bar\tau z_j^2 + \tau z_{j+1}^2}
=
\bar\tau' I [z_j^M] + \tau' I [z_{j+1}^M]
\, .
\end{align}
So, up to a redefinition of the proper time,
\begin{align*}
\tau' = \frac{\tau}{\tau + \bar\tau z_j^2/z_{j+1}^2}
\, ,
\end{align*}
the path retains its form. Thus classically, the $D$-dimensional Wilson loop is invariant under the conformal contour deformations.

\subsection{Lagrangian insertions}

Since off-shell amplitudes rapidly become complicated and extremely laborious to calculate with increasing loop order, we prefer to deal with their integrands. This inevitably forces 
us to use the formalism of operator insertions familiar from studies of renormalization of composite operators \cite{Kluberg-Stern:1974iel}, which was recently popularized within the 
context of correlation functions starting from \cite{Howe:1998zi}. The usefulness of the method of Lagrangian insertion for the Wilson loop is practically twofold. First, as we just 
said, it allows one to define the notion of the integrand for the bosonic Wilson loop starting already from the one loop order. Unfortunately, this procedure does not offer any computational 
simplifications in perturbative studies on par to the correlator story. Notwithstanding, its very existence is complemented by yet another feature of paramount importance: the ability to 
analytically continue and vary the $(2 \varepsilon)$-regulator dimensions independently of the off-shellness condition for external kinematical variables. The reason for this is once
one imposes the null conditions \re{LCcondition}, the integrand becomes a function of two independent regulator variables, the off-shellness of the perimeter and dimensionality of the 
`internal' space of the Wilson loop surface. It is this property that will let us to continuously navigate between the off-shell case in 4D and the on-shell one in the dimensionally regularized 
theory.

Starting with the perturbative expansion of the Wilson loop expectation value in YM coupling
\begin{align}
\label{WLPT}
\vev{W_N} = 1 + g^2_{\scriptscriptstyle\rm YM} \vev{W_N}^{(1)} + g^4_{\scriptscriptstyle\rm YM}  \vev{W_N}^{(2)} + \dots \,,
\end{align}
one finds individual coefficients in the series by differentiating both sides of this equation with respect to $g^2_{\scriptscriptstyle\rm YM}$ sufficient number of times and setting it afterwards 
to zero. In the spirit of the computation of correlation functions \cite{Howe:1998zi,Arutyunov:2003ad,Alday:2010zy}, we find then at one and two loops, 
\begin{align}
\label{WLVEVinsertions}
\vev{W_N}^{(n)}  
&
= \left. \frac{1}{n! g_{\scriptscriptstyle\rm YM}^{2 n}}
\int [dA] \e^{i S} W_N \Delta^{(n)} \right|_{g_{\scriptscriptstyle\rm YM} \to 0}
\, ,
\end{align}
where we rescaled the gauge field back by the YM coupling, i.e., $A_M \to g_{\scriptscriptstyle\rm YM} A_M$ to regain standard normalization of interaction terms in the gauge 
action for easier counting of therms contributing to the same order of perturbation theory. To the lowest two perturbative orders, the differential operator insertions 
\begin{align}
\label{OffShellWilson}
\Delta^{(1)}
&=
i \int d^D z_0 \mathcal{L}^\prime (z_0)
\, , \\
\Delta^{(2)}
&=
\left(
i\int d^D z_0\mathcal{L}^\prime (z_0)
\right)^2
+
i \int d^D z_0 \mathcal{L}^{\prime\prime} (z_0)
\, ,
\end{align}
are determined by the derivatives of the Lagrangian \re{TruncLag} with respect to $g^2_{\scriptscriptstyle\rm YM}$,
\begin{align}
\label{LagrangianInsertion}
\mathcal{L}^\prime 
&
= {\rm tr} \,
\bigg(
\ft12
F_{MN} F^{MN}
-
\ft{1}{\sqrt{2}} g_{\rm YM}^3
\lambda^{\alpha A}
[\bar\phi_{AB}, \lambda_\alpha^B]
+
\ft{1}{\sqrt{2}} g_{\rm YM}^3
\bar\lambda_{\dot\alpha A}
[\phi^{AB}, \bar\lambda^{\dot\alpha}_B]
+
\ft{1}8 g_{\rm YM}^2
[\phi^{AB}, \phi^{CD}] [\bar\phi_{AB}, \bar\phi_{CD}]
\bigg)
, \nonumber\\
\mathcal{L}^{\prime\prime} 
&
= {\rm tr} \,
\bigg(
- 
F_{MN} F^{MN}
+
\ft{1}{2\sqrt{2}} g_{\rm YM}^3
\lambda^{\alpha A}
[\bar\phi_{AB}, \lambda_\alpha^B]
-
\ft{1}{2\sqrt{2}} g_{\rm YM}^3
\bar\lambda_{\dot\alpha A}
[\phi^{AB}, \bar\lambda^{\dot\alpha}_B]
\bigg)
. 
\end{align}
Here, compared to Eq.\ \re{TruncLag}, the gluon fields are rescaled by the power of the YM coupling, the fermion and scalar fields are restored, but
the gauge fixing and ghost terms being dropped by virtue of their property to yield vanishing expectation values when inserted with gauge-invariant (Wilson
loop) operators. 

By considering the integrands of the Wilson loop
\begin{align}
\int [dA] \e^{i S} W_N \mathcal{L}^\prime (z_0) \dots \mathcal{L}^\prime (z_0')
\, ,
\end{align}
etc., the fields along the Wilson loop's perimeter can Wick contract among themselves only through the Lagrangian insertions or through the interaction terms 
brought down from the action in the Boltzmann weight factor. In this manner, the Wilson loop is regarded as a normal-ordered operator\footnote{The 
Lagrangian itself is normal ordered as well.}. Therefore, there are two types of propagators involved, the bulk-to-boundary and the bulk-to-bulk. The former propagate 
fields from the perimeter to the interior interaction points of the loop, while the latter only connect interior points. This allows us to separate short-distance 
regulators for the boundary and the bulk when we address quantum mechanical effects. 

The gauge propagator between the boundary and the bulk point $z_0$ reads
\begin{align}
\label{GluonProp}
\vev{ A^a_M (z_0) A^b_N (z_{jj+1}^{\tau_j})} 
= 
-
\frac{1}{4 \pi^{D/2}} \eta_{MN} \delta^{ab} \frac{\Gamma (\frac{D}{2} - 1)}{[- (z_0 - z^{\tau_j}_{jj+1})^2 + i 0]^{D/2-1}}
\, ,
\end{align}
in the Feynman gauge. We set the conventional dimensional regulator scale to unity $\mu = 1$, so as not to pollute formulas, however, we will recover it when it becomes important 
for our arguments. The resulting integrand is a function of the space-time dimension $D$ of the Lagrangian insertions and of the external kinematics, in particular, the off-shellness 
$x_{jj+1}^2 = m^2$. We consider it as an independent function of both. Since the short-distance divergences around the perimeter are now regularized by $m$ and we can safely send 
$\varepsilon \to 0$ in the bulk\footnote{Of course, when closed loops arise in the interior of the Wilson loop at higher orders, they will induce unregularized singularities. We will return to 
this later on when discussing two-loop effects.}. Thus, at this step we project out the overlap between the internal and external coordinates in extra dimensions. It is important that this
dimensional reduction of the interior down to Minkowski space is taken at the very end of all field Lorentz contraction, use of equations-of-motion etc. With this understanding of the 
integrand, the invariant distance in the bulk-to-boundary propagator reduces to 
\begin{align}
\label{DdimDistances}
(x_0 - z^{\tau_j}_{jj+1})^2
=
(x_0 - x^{\tau_j}_{jj+1})^2 - (y^{\tau_j}_{jj+1})^2 
= 
(x_0 - x^{\tau_j}_{jj+1})^2 + \tau_j \bar\tau_j x_{jj+1}^2
\, ,
\end{align}
where the light-cone condition on the external kinematics \re{LCcondition} was used simultaneously with the null conditions \re{NullYs} on the individual $y_j$ variables. The resulting 
Wilson loop in the off-shell kinematics and the four-dimensional interior can seen in the panel $(b)$ of Fig.\ \ref{WilsonLoopPic}: trajectories of the probes sourcing the gauge flux cease to be 
straight but rather develop a pull into the auxiliary kinematical directions encoded by a unit vector $n_j$ orthogonal to the Minkowski plane, i.e., $z_{jj + 1}^{\tau_j} \to \bar\tau_j x_{j} + 
\tau_j x_{j + 1} + m \sqrt{\tau_j \bar\tau_j} n_j$. As we will see in the next subsection, this setup yields an exact one-loop match between the integrand of the off-shell four-gluon scattering 
amplitude \re{OffShellAmplitude} and the integrand of the Wilson loop constructed in this manner. 

Finally, the bulk-to-bulk propagator has the form identical to \re{GluonProp} except that the intervals are all four-dimensional (after dimensional reduction). Below, we will need these only in 
polarization tensors since in all other circumstances we will be able to integrate these out in the path integral before any calculations are performed at two loops.

\section{One-loop test}
\label{SectionOneLoop}

\begin{figure}[t]
\begin{center}
\mbox{
\begin{picture}(0,120)(200,0)
\put(0,0){\insertfig{14}{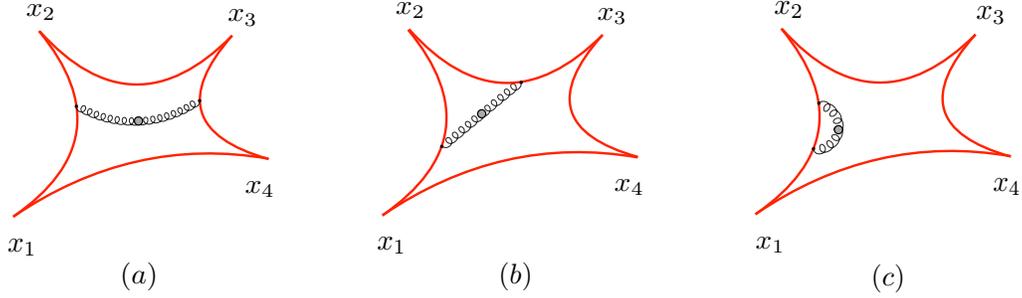}}
\end{picture}
}
\end{center}
\caption{\label{OneLoopPic}  One-loop graphs for the integrand of the off-shell four-site Wilson loop. The grey blob stands for the
Lagrangian insertion \re{LagrangianInsertion}.}
\end{figure}

The first order of business is to demonstrate the equivalence of the just proposed Wilson loop along with its calculational procedure to the result quoted in Eq.\ \re{OffShellAmplitude}. 
The contributing one-loop graphs are shown in Fig.\ \ref{OneLoopPic} (and their cyclic permutations). Their integrands possess the structure\footnote{We define the free expectation 
value $\vev{ \dots }^{(0)} \equiv \int [dA] \exp(i S_0) \dots$ as a path integral averaged with the Boltzmann weight of the free gluon action $S_0 = - \frac{1}{2} \int d^D x_0 (\partial_M 
A_N^a)^2$ in Feynman gauge $\xi = 1$.}
\begin{align}
\label{GegericOneLoop}
\VEV{\frac{1}{N_c} P \tr \left[ w_{[j+1,j]}w_{[j'+1,j']}\mathcal{L}^\prime (x_0) \right]}^{(0)}
=
- \frac{C_F}{4 \pi^4}
\int_0^1 d\tau_j \int_0^1 d\tau_{j'} 
\frac{n (x_0; z_j, z_{j + 1}; z_{j'}, z_{j' + 1} )}{\big(x_0 - z^{\tau_j}_{j j + 1}\big)^4 \big(x_0 - z^{\tau_j'}_{j' j' + 1} \big)^4}
\, .
\end{align}
and are proportional to the quadratic Casimir in the fundamental representation of the SU($N_c$) group $C_F = (N_c^2-  1)/(2N_c)$. The numerator, while superficially appearing to have 
a nontrivial dependence on the proper times $\tau_{j, j'}$, is in fact independent of them
\begin{align}
\label{OneLoopNumerator}
n (x_0; z_j, z_{j + 1}; z_{j'}, z_{j' + 1} )
=
z_{j j +1} \cdot z_{j' j' +1} \,  (x_0 - z_j) \cdot (x_0 - z_{j'})
-
z_{j j +1} \cdot (x_0 - z_{j'}) \,  z_{j' j' +1} \cdot (x_0 - z_{j})
\, ,
\end{align}
due to the transverse nature of the Abelian portion of the field strength tensor. This is the first harbinger of successful comparison to the amplitude \re{OffShellAmplitude}: correlation
functions computed with the Lagrangian insertion procedure in the pair-wise light-cone limits possess the very same numerators on a diagram-by-diagram basis. Notice here that
all Lorentz contractions were performed in $D$-dimensions before projecting out $2\varepsilon$-directions of the bulk points since otherwise one would be left only with 
four-dimensional inner products of coordinates. The second step we need to perform is to convert and match the proper-time integrals to the expected denominators of the cross 
integral $g_{1234}$ in Eq.\ \re{CrossIntegrals}. This follows immediately upon rearranging the denominators in Eq.\ \re{GegericOneLoop} by
\begin{align}
\big(x_0 - z^{\tau_j}_{j j+1}\big)^2 = \bar\tau_j x_{0 j}^2 + \tau_j x_{0 j + 1}^2
\, ,
\end{align}
with any explicit mentioning of the off-shellness lost in translation. The remaining $\tau$-integrals immediately produce products of the free scalar propagators $1/x_{0 j}^2$. Now, all that is left
to do is to show that the sum of all graphs, displayed in Fig.\ \ref{OneLoopPic} and their inequivalent permutations, add up to $z_{13}^2 z_{24}^2 g_{1234}$. This is almost warranted by the
equivalence of the numerators between the integrands of the Wilson loop and correlation functions alluded to above and can be verified by rearrangement of numerators for the three  
graphs in Fig.\ \ref{OneLoopPic}. Namely, one finds
\begin{align}
\label{EqExchange}
n (x_0; z_1, z_2; z_3, z_4 )
&
= \frac{1}{4}
\left[
z_{13}^2 z_{24}^2 - z_{13}^2 [x_{04}^2 + x_{02}^2] - z_{24}^2 [x_{01}^2 + x_{03}^2]
-
[x_{01}^2 - x_{02}^2] [x_{03}^2 -x_{04}^2 ]
\right]
\, , \\
\label{EqCusp}
n (x_0; z_1, z_2; z_2, z_3 )
&
= \frac{1}{4}
\left[
2 z_{13}^2 x_{02}^2
-
[x_{01}^2 - x_{02}^2] [x_{02}^2 -x_{03}^2]
\right]
\, , \\
\label{EqSelf}
n (x_0; z_1, z_2; z_1, z_2 )
&
= 
- \frac{1}{2}
[x_{01}^2 - x_{02}^2] ^2
\, , 
\end{align}
for the diagrams in $(a)$, $(b)$ and $(c)$, respectively. The cancellation mechanism is then self-explanatory, i.e., the second and third terms in Eq.\ \re{EqExchange}
(and its analogue with the gluon insertion stretching between the other opposite sides) cancel against the first term in the vertex contribution \re{EqCusp} and its cyclic 
permutations, while the fourth term in \re{EqExchange} together with the last one in \re{EqExchange} conspire to cancel the self-energy contributions in Eq.\ \re{EqSelf}
and its permutations. This leaves us just the (double of the) first term in \re{EqExchange}, hence confirming the agreement between the one-loop off-shell Wilson loop 
and the one-loop gluon amplitude \re{OffShellAmplitude} in the planar approximation $C_F \to N_c/2$. The one-loop integrand is explicitly conformal invariant inheriting
the inversion symmetry \re{InvarianceContour}. This gives the first (and, unfortunately, the last) evidence in favor of our proposal.

\section{Off-shell to on-shell}
\label{OffOnShellSection}

In this section, we relax the condition for internal integrations to be four-dimensional.  While massless one-loop calculations can be done on the back of an envelop, massive 
calculations in dimensionally regularized (or not) theory are far from being trivial. The motivation for this is to study the dependence of the Wilson loop expectation value 
simultaneously both on the off-shellness and dimensional regulator parameters and thus to be able to interpolate between the off- and on-shell regimes. When the segments' 
`virtualities' tend to zero, the Wilson loop becomes light-like and suffers from well-known ultraviolet divergences: $m^2 = 0$ no longer cuts-off short-distance behavior of gluons in
the vicinity of the cusps. To tame them properly, we have to analytically continue $\varepsilon$ in $D = 4 + 2 \varepsilon$ in Eq.\ \re{OffShellWilson} to negative values $\varepsilon
\equiv -\epsilon <0$. This is how we will understand the dimensionality of the internal integration in this section. The distances are however taken to be same as  they arise from 
the dimensional reduction, i.e., see  Eq.\ \re{DdimDistances}.

When the space-dimension $D$ is kept away from four, the massive cancellation observed between various graphs observed at the end of the previous section no longer holds except 
for the self-energy-like terms, i.e., the last contributions in Eq.\ \re{EqExchange} and \re{EqCusp} and Eq.\ \re{EqSelf} itself, which are total derivatives as we show in the Appendix A. 
So the exchange and the vertex graphs have to be computed from scratch. To cast the emerging parametric integrals in the as most eye-pleasing form as possible, we use a somewhat 
unorthodox version of the Feynman parametrization to join the two denominators in Eq.\ \re{GegericOneLoop} promoted to their $D$-dimensional versions \re{GluonProp},
\begin{align*}
\frac{\Gamma (\nu_1) \Gamma (\nu_2)}{A_1^{\nu_1} A_2^{\nu_2}}
=
\tau_{12}
\int_{\tau_2}^{\tau_1} d \sigma (\tau_1 - \sigma)^{\nu_1 - 1} (\sigma - \tau_2)^{\nu_2 - 1}\frac{\Gamma (\nu_1 + \nu_2)}{[(\tau_1 - \sigma) A_2 + (\sigma - \tau_2) A_1]^{\nu_1 + \nu_2}}
\, ,
\end{align*}
with the adopted short-hand notation $\tau_{12} \equiv \tau_1 - \tau_2$.
The right-hand side is independent of the choice for $\tau$'s and they can be selected at will. To achieve the above simplification they are taken to be $\tau_1 = \tau_j$ and
$\tau_2 = \tau_{j'}$ for $A_1 = - (x_0 - z^{\tau_j}_{j j + 1})^2$ and $A_2 = - (x_0 - z^{\tau_j'}_{j' j' + 1})^2$. Performing the integral over the insertion point $z_0$,
we find\footnote{We do not display the self-energy-like contributions in $(a)$ and $(b)$ diagrams since their sum cancels in $D$-dimensions against the self-energy graph $(c)$ in the 
same fashion as in the four dimensions as we pointed out at the end of the previous section. Their structure is made explicit in the Appendix A which makes this statement obvious.} 
\begin{align}
\vev{ W_{4} }^{(1)}_{{\rm Fig.} \ref{OneLoopPic} (a)} 
&
= 
-
\frac{C_F}{16 \pi^{D/2}}
\left\langle\!\!\!\left\langle
\left[
z_{13}^2 z_{24}^2 
- 
z_{13}^2 [x_{24}^2 \sigma + 2 m^2 \bar\sigma]
- 
z_{24}^2[x_{13}^2 \sigma + 2 m^2 \bar\sigma]
\right]
\frac{\Gamma (\frac{D}{2})}{\mathcal{D}_{x_{13},  x_{24}}^{D/2}}
\right.\right.
\nonumber\\
&
\qquad\qquad
\qquad\qquad
\qquad\quad\!
\qquad
\left.\left.
+\, 2 \frac{z_{13}^2 + z_{24}^2}{\tau_{12}^2}
\frac{\Gamma (\frac{D}{2}-1)}{\mathcal{D}_{x_{13},  x_{24}}^{D/2-1}}
\right\rangle\!\!\!\right\rangle
\, , \nonumber\\
\vev{ W_{4} }^{(1)}_{{\rm Fig.} \ref{OneLoopPic} (b)} 
&
= 
- \frac{C_F}{16 \pi^{D/2}}
\left\langle\!\!\!\left\langle
2 z_{13}^2 m^2 \bar\sigma
\frac{\Gamma (\frac{D}{2})}{\mathcal{D}_{x_{13},  0}^{D/2}}
-
\, 2 \frac{z_{13}^2}{\tau_{12}^2}
\frac{\Gamma (\frac{D}{2}-1)}{\mathcal{D}_{x_{13},  0}^{D/2-1}}
\right\rangle\!\!\!\right\rangle
\, , 
\end{align}
for the respective contributing graphs $(a)$, $(b)$ to the Wilson loop. Notice that the integrand of the vertex diagram is related to the one of the exchange graph by sending 
$4 \to 2$ and flipping the overall sign. Here the double angle brackets stand for the three-fold parametric integral
\begin{align}
\langle\langle \dots \rangle\rangle
\equiv 
\int_0^1 d\tau_1 \int_0^1 d \tau_2 \, \tau_{12} \int_{\tau_2}^{\tau_1} d \sigma [(\tau_1 - \sigma) (\sigma - \tau_2)]^{D/2-1}
\dots
\, ,
\end{align}
and the denominator being
\begin{align}
\label{ExchangeDenominator}
\mathcal{D}_{x_{13},  x_{24}} \equiv (\tau_1 - \sigma) (\sigma - \tau_2) \, [ - \bar{\tau}_1 \bar{\tau}_2 x_{13}^2 - \tau_1 \tau_2 x_{24}^2 ] - \sigma\bar\sigma \tau_{12}^2 m^2
\, .
\end{align}

The above parametric integrals are not immediately expressible in terms of known special functions. However, it is relatively straightforward to derive Mellin-Barnes representations
for them. Details of calculations involved are relegated to the Appendix A. Here we merely cite the result for the main structure $z_{13}^2 z_{24}^2$ which survives the four-dimensional
limit. We found the following three-fold contour integral for it
\begin{align}
\label{MBtripple}
\left\langle\!\!\!\left\langle
\frac{\Gamma (\frac{D}{2})}{\mathcal{D}_{x_{13},  x_{24}}^{D/2}}
\right\rangle\!\!\!\right\rangle
=
\frac{(-m^2)^{-D/2}}{\Gamma^2 \left(2-\frac{D}{2} \right)}
\int \prod_{j = 1}^3\frac{d z_j}{2 \pi i} \Gamma (-z_j)
\left(
\frac{x_{13}^2}{m^2}
\right)^{z_1}
\left(
\frac{x_{24}^2}{m^2}
\right)^{z_2}
\phantom{\hspace{100pt}}
\end{align}
\vspace{-16pt}
\footnotesize
\begin{align*}
\times
\frac{
\Gamma (-z_3) \Gamma (z_1+z_3+1) \Gamma (z_2+z_3+1) \Gamma \left(1-\frac{D}{2}-z_1-z_3\right) 
\Gamma \left(1-\frac{D}{2}-z_2-z_3\right) \Gamma \left(\frac{D}{2}+z_1+z_2+z_3\right)^2
}{
\Gamma \left(\frac{D}{2}+z_1+z_2\right)}
\, . 
\end{align*}
\normalsize
Other contributions to the exchange graph are unfortunately not as simple: they admit a four-fold Mellin-Barnes representation which can be found in Eq.\ \re{MBfourfold}.
Finally, the vertex integral is given in Eq.\ \re{VertexResult}. The complete result for the off-shell dimensionally regularized one-loop Wilson is then obtained by adding one 
more exchange contribution and corrections to other three cusps of the four-site Wilson loop, namely,
\begin{align}
\vev{W_{4}}^{(1)} 
=
\sum_{{\rm cyclic} (1234)} 
\left[
\frac{1}{2}
\vev{ W_{4} }^{(1)}_{{\rm Fig.} \ref{OneLoopPic} (a)} 
+
\vev{ W_{4} }^{(1)}_{{\rm Fig.} \ref{OneLoopPic} (b)} 
\right]
\, .
\end{align}
with the factor of $1/2$ eliminating the double counting of the exchanged graphs in the sum.

With these generic expressions in hand, we would like to extract anomalous dimensions governing the Sudakov behavior in the off- and on-shell regimes. These can easily be 
deduced by studying the following two limits: the four-dimensional vanishing off-shellness and the on-shell dimensionally regularized case. These respective asymptotic expressions 
read\footnote{The pole contribution is ${\rm Poles} = - 2 \varepsilon^{-2} + \varepsilon^{-1} \log(\mu^4/x_{13}^2 x_{24}^2)$.}
\begin{align}
\label{OffShellWL1loop}
\log \vev{W_4}_{D = 4, m^2 \to 0}
= 
&
- 4 g^2
\left[
\log^2 \frac{m^2}{\sqrt{x_{13}^2 x_{24}^2}} + \frac{1}{2} \zeta_2
\right]
\, , \\
\log \vev{W_4}_{D \neq 4, m^2 = 0}
=
&
- 2 g^2
\left[
\log \frac{\mu^{2}}{x_{13}^2} \log \frac{\mu^{2}}{x_{24}^2} - 2 \zeta_2
+
{\rm Poles}
\right]
\, ,
\end{align}
as $x_{13}^2, x_{24}^2 \to \infty$. In the second formula, we restored the renormalization scale $\mu$ dependence and absorbed in it certain accompanying transcendental 
constants along the way. We observe indeed the well-known `factor of two difference' between the Sudakov asymptotics in the off-shell \cite{Sudakov:1954sw} and on-shell 
regimes \cite{Jackiw:1968zz}, which is attributed to the non-vanishing contribution from ultra-soft integration in addition to the soft regime, present in both cases, as explained 
in Ref.\ \cite{Fishbane:1971jz} (see, in particular, Appendix  A there), \cite{Mueller:1981sg} (see section 12.1 there) and Ref.\ \cite{Korchemsky:1988hd}. These one-loop 
results are in accord with considerations of the on-shell \cite{Bern:2005iz} and off-shell four-leg one-loop amplitude in Refs.\ \cite{Drummond:2007aua} and 
\cite{Caron-Huot:2021usw}. However, the off-shell Wilson loop \re{OffShellAmplitude} starts to deviate from the latter two starting from two loops.

\section{Two loops}
\label{2LoopsSection}

Encouraged by the findings of the previous sections, we move on to the two-loop order. An off-shell (or rather massive) two-loop calculation is a highly nontrivial endeavor. 
Therefore, we will presently address only the small-$m$ limit of the off-shell Wilson loop with four sites. We aim to achieve two goals, first, we need to confirm that the 
perimeter regularization is consistent with the Wilson loop exponentiation and, two, if this is the case, what the corresponding function of the coupling is that governs this 
Sudakov behavior. So all we need to do is to track the logarithms of the soft scale, which can be made dimensionless either by $x_{13}^2$ or $x_{24}^2$. Therefore, for our 
present needs, it suffices to impose the symmetric-point condition $x_{13}^2 = x_{24}^2$ and thus introduce a single scaling variable $\Delta^2 \equiv x_{13}^2/m^2$ driving 
the asymptotic behavior of the problem\footnote{This kinematic constraint implies that we will not be able to trace the fate of the conformal symmetry at two loops.}.

Next, we would like to simplify the form of the Lagrangian insertion making use of the gauge invariance of the Wilson loop. Namely, we choose a gauge-fixed form of the Lagrangian
for this purposes. This merely entails a substitution of the first term in $\mathcal{L}'$ by the integrand of Eq.\ \re{TruncLag}. In the Feynman gauge, which we use throughout the 
calculation, the kinetic term of the YM action becomes particularly simple
\begin{align}
\label{FGgluonaction}
S_0 = - \int d^D x_0 \tr (\partial_M A_N)^2
\, .
\end{align}
It is merely a free-field equation of motion. This will allow us to easily integrate it out in the path integral when it appears along with interaction terms from the action. Recall that we 
are not able to do this when $S_0$ is connecting any two sites of the loop by free Feynman propagators \re{GluonProp} due to normal ordering constraints on $W_N$. To be very
specific this discussion implies the following
\begin{align}
\label{IntegratingOUTs0}
\vev{{\rm N}[ {\dots} O_1] S_0 {\rm N}[O_2 {\dots} ]}^{(0)} = i \vev{{\rm N}[\dots O_1] {\rm N}[ O_2 \dots]}^{(0)}
\, , \quad
\vev{{\rm N}[\dots O_1 O_2] S_0}^{(0)} \neq i \vev{{\rm N}[\dots O_1 O_2 ]}^{(0)}
\, ,
\end{align}
where we temporarily made the normal ordering explicit.

Now the procedure for the calculation of $\vev{W_4^{(2)}}$ from Eq.\ \re{WLVEVinsertions} is straightforward. The expansion of the Wilson loop links\footnote{Recall that we 
rescaled $A$ there by $g_{\scriptscriptstyle\rm YM}$ to have the usual counting of its powers to match a given quantum order.} \re{WLsegment} in YM coupling is equivalent 
to the number of sites (same site can appear more than once) involved in the interaction,
\begin{align}
\label{CiteExpansion}
W_4 = 1 + g_{\scriptscriptstyle\rm YM}^2  w_2 + g_{\scriptscriptstyle\rm YM}^3 w_3 + g_{\scriptscriptstyle\rm YM}^4 w_4 + \dots
\, ,
\end{align}
where
\begin{align}
w_k = \frac{1}{N_c} \tr P \left( i \int d z_M  A^M (z) \right)^k
\, ,
\end{align}
and the ellipses stand for higher order terms which are irrelevant for the current two-loop analysis. Substituting the above series into Eq.\ \re{WLVEVinsertions} for $n=2$,
we naturally introduce the following nomenclature for contributions in the Wilson loop average
\begin{align}
\vev{W_4}^{(2)} = \mathcal{W}_2^{(2)} + \mathcal{W}_3^{(2)} + \mathcal{W}_4^{(2)}
\end{align}
corresponding to the two, three and four Wilson links involved in the interaction, respectively,
\begin{align}
\mathcal{W}_k^{(2)} 
= \left. \frac{1}{2} 
g_{\scriptscriptstyle\rm YM}^{k-4}
\int [dA] \e^{i S} w_k \Delta^{(2)} \right|_{g_{\scriptscriptstyle\rm YM} \to 0}
\, .
\end{align}

Before we proceed with explicit calculations at two loops, first, we return back to the one-loop contributions to demonstrate a formalism which will be easily generalizable to 
higher loop orders in the analysis of the asymptotic $m \to 0$ behavior of the Wilson loop.

\begin{figure}[t]
\begin{center}
\mbox{
\begin{picture}(0,250)(210,0)
\put(0,0){\insertfig{15}{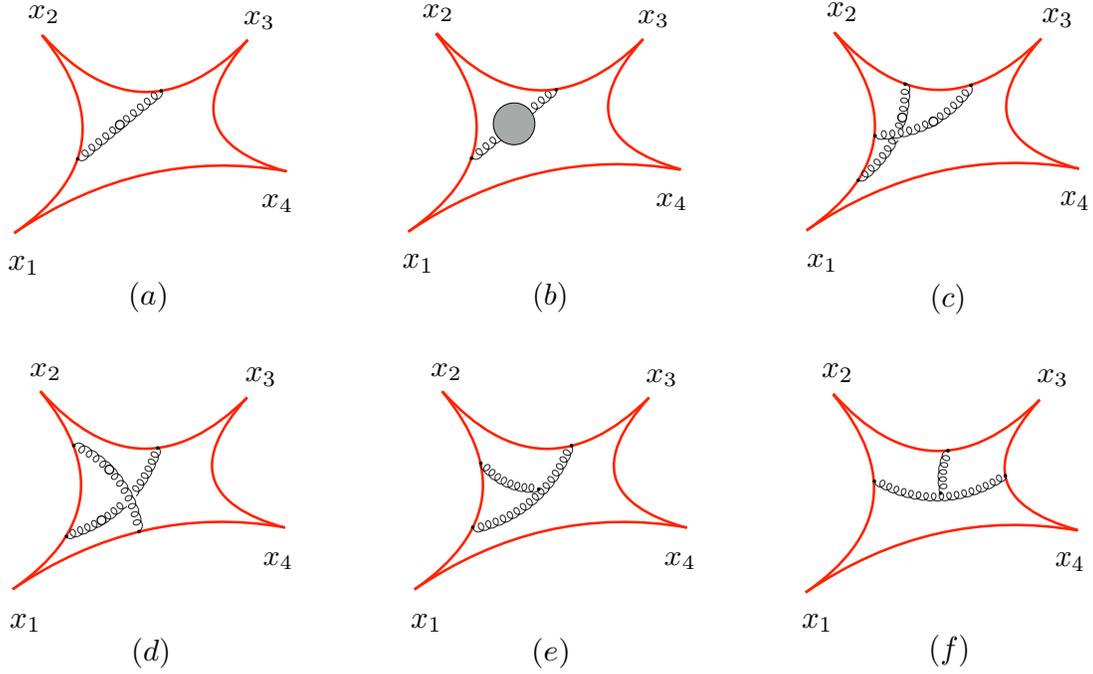}}
\end{picture}
}
\end{center}
\caption{\label{12loopWLpic} One- and two-loop Feynman graphs possessing at least double logarithmic dependence on the off-shellness.
The large grey blob stands for the virtual loops of gluons, ghosts, fermions and scalars, while the small empty one is the insertion of the Feynman
gauge YM action \re{FGgluonaction}. The multiplicity factors of the displayed graphs are (from left top to right bottom) $m_\alpha = \{4, 4, 4, 4, 8, 4\}$.}
\end{figure}

\subsection{Back to one loop}

At one loop, we only need to keep the two-gluon term $w_2$ in the expansion of the path-ordered exponential such that
\begin{align}
\vev{W_4}^{(1)} 
=
- i \VEV{w_2 S_0 }^{(0)}
\end{align}
The only graph which induces the double logarithmic scaling is the one where a gluon dresses up a single cusp. One representative example is given Fig.\ \ref{12loopWLpic} $(a)$
and its coordinate space integral, after dimensional reduction of the interior to four-dimensions, reads\footnote{We drop the $2\varepsilon$-portion of the $D$-dimensional 
scalar products in the numerator, c.f., Eq.\ \re{EqExchange}, since these cannot affect the Sudakov scaling we are after.}
\begin{align}
\vev{W_4}^{(1)}_{{\rm Fig.} \ref{12loopWLpic} (a)}
= - \frac{i C_F}{(2 \pi)^4} 
( x_{12} \cdot x_{23} )
\int_0^1 ds \int_0^1 dt \int d^4 x_0 \, 
\partial_0 D (x_0; x_{12}^{s}) \cdot  \partial_0 D (x_0; x_{23}^{t}) 
\, , \nonumber
\end{align}
in terms of the stripped bulk-to-boundary propagators
\begin{align}
D (x_0; x_{jj+1}^{s} ) = 1/[(x_0 - x_{jj+1}^{s} )^2 + m^2 s \bar{s} ]
\, .
\end{align}
The four-dimensional portion of the path is parametrized by a straight line, i.e., $x_{jj+1}^{s} = \bar{s} x_j + s x_{j+1}$ and its curvilinearity is encoded in the mass term.
The coordinate-space integral can be evaluated using conventional Feynman parametrization and the following basic integral
\begin{align}
\frac{i}{\pi^{D/2}}
\int d^D x_0 \frac{\Gamma(\alpha)}{[ - x_0^2 - L]^\alpha} =  \frac{\Gamma(\alpha - D/2)}{[- L]^{\alpha - D/2}} 
\, .
\end{align}
In this manner, we find
\begin{align}
\vev{W_4}^{(1)}_{{\rm Fig.} \ref{12loopWLpic} (a)}
=
&-
\frac{C_F}{(2 \pi)^2} \int_0^1 ds \int_0^1 d t \int_0^1 d\alpha_1 d \alpha_2 \delta (\alpha_1 + \alpha_2 - 1)
{\rm Int}^{(1)}_{[32],[21]} (x_{12}^{s}, x_{23}^{t}; \alpha_1, \alpha_2)
\, ,
\end{align}
with the one-loop parametric integrand being
\begin{align}
\label{1loopIntegrand}
{\rm Int}^{(1)}_{[32],[21]} (x_{12}^{s}, x_{23}^{t}; \alpha_1, \alpha_2)
=
\frac{ ( x_{12} \cdot x_{23} ) }{L_{[32],[21]}} 
+
m^2 \left( s \bar{s}/ \alpha_2 + t \bar{t}/\alpha_1 \right) \frac{ ( x_{12} \cdot x_{23} ) }{L_{[32],[21]}^2} 
\, ,
\end{align}
where the denominator is
\begin{align}
L_{[32],[21]} = (x_{12}^{s} - x_{23}^{t})^2 + \frac{s \bar{s}}{\alpha_2} m^2 + \frac{t \bar{t}}{\alpha_1} m^2 
\, .
\end{align}
What is the most efficient way to extract the $m^2 \to 0$ asymptotic behavior of this integral that can be easily generalized to higher loops? We answer this 
question in the next two sections\footnote{Running ahead of our presentation, let us point out that we will employ the method of the Mellin-Barnes representation. 
In recent years the latter has lost its apex position on the stage of Feynman integrals being superseded by a more powerful method of differential equations.
However, in our situation, the differential equations can hardly be applied so that we will rely on the good old Mellin-Barnes technique.}.

\subsection{Method of regions} 

To evaluate the leading and next-to-leading logarithms of the parameter $\Delta$ in the limit $\Delta\to \infty$, we apply the strategy of expansion by regions which 
was introduced in the case of threshold expansions of Feynman integrals in \cite{Beneke:1997zp} (see also \cite{Smirnov:2002pj} and \cite{Smirnov:2012gma}) and 
formulated for momentum integrals in the language of the Feynman-parameter representation in \cite{Smirnov:1999bza}. According to this strategy, the expansion of a given 
Feynman integral is given by a sum over so-called {\sl regions}. In a sense, it is akin to (but more general than) the well-known Wilson operator product expansion on the 
diagrammatic level.
 
Typically, the task of determining all contributing regions governing a given asymptotics is a highly tedious task. An algorithmic way to reveal them was found 
in \cite{Pak:2010pt}. It is based on the geometry of polytopes connected with basic functions determining the Feynman-parameter integrand. This was a very important step both 
for theoretical as well as practical reasons. First, it allowed one to formulate expansion by regions in an unambiguous mathematical framework. Second, a first public code 
{\tt asy} was made available by Ref.\ \cite{Pak:2010pt}. The latter was further improved in \cite{Jantzen:2012mw} and successfully applied to various problems involving 
Feynman integrals.  The most recent version of the code {\tt asy2.1.m}  is available from the {\tt FIESTA5} distribution package \cite{Smirnov:2021rhf}.
 
The expansion by regions has the status of an experimental mathematics. However, currently there are no known counterexamples where this strategy was shown to fail. An interested
reader can find a discussion of the problem with the mathematical justification of this strategy in Ref.\ \cite{Semenova:2018cwy} as well as in the comprehensive review
\cite{Smirnov:2021EbR}. 
 
As we have already seen in the previous section at one loop, and as it will be even more obvious below, all graphs are expressed in terms of parametric integrals over 
multidimensional unit cube where the number of integration variables is less or equal to six. To be able to literally apply the code {\tt asy} to the problem at hand, we need to 
map this hypercube to $\mathbb{R}_+^n$. This is easily accomplished by the following transformation of variables
\begin{align}
\label{UnitCubeToRn}
\alpha_j = \frac{x_j}{1+x_j}
\, ,
\end{align}
for all proper times ($s,t$, etc) and Feynman $\alpha$-parameters. The integration now runs over the half infinite interval $0 \leq x_j < \infty$. For each Feynman graph, we obtain an 
integral of a rational function of ${\delta,x_1,x_2,\ldots}$, where we inverted the scaling variable and defined $\delta\equiv 1/\Delta$. The denominator in the integrand is a product
of simple factors $(1+x_j)$ in positive integer powers coming from the Jacobian as well as a 
polynomial $P(\delta; x_1,x_2,\ldots)$ which is unapologetically cumbersome since it 
inherits the structure of the Feynman-joined bulk-to-boundary propagators. Let us emphasize that, in our calculation, we encounter only polynomials with positive coefficients, so that there are no complications connected with negative terms, like in the case of the threshold expansion~\cite{Jantzen:2012mw}.

As in the case of well-known application of the Wilson operator product expansion, even if one starts with a finite observable (say, a cross section for deep-inelastic scattering),
individual terms in its twist expansion develop singularities due to approximations involved and thus require a regularization. At the level of dimensionally regularized Feynman 
graphs, when one expands them at the level of the integrand for large/small values of external kinematical variables, these develop extra poles in the parameter of dimensional 
regularization $\epsilon = (4-D)/2$. These extra poles have to cancel at every order of the expansion and thus provide a crucial self-consistency check on the correctness of the
calculation: the sum of pole parts of the expanded integral in $\epsilon$ equals the pole part of the initial integral. 

However, even if the initial integral is convergent in four-dimensions, we are forced to introduce a regularization in order for individual terms in the expansion by regions 
to be well defined as $\delta\to 0$. This is easily accomplished by introducing an analytic regularization for the integrand's denominator. Since the $\delta$-dependence
only enters in the polynomial $P$ introduced earlier, it suffices to just shift its power $n$ by some parameter $\eta$,  i.e., $P^{n} \to P^{n - \eta}$.

Let us now give a few hints of the code use. Suppose that the denominator of a four-fold parametric integral has the form $(1 + x_1)^2(1 + x_2)^3(1+x_4) P^{2} 
(\delta; x_1,x_2, x_3,x_4)$. Then to reveal all regions of the parameter space, we turn to {\tt asy}  and run  
\begin{align*}
\begin{tt}
r=WilsonExpand[(1 + x_1)(1 + x_2)(1+x_4),P, \{x_1,x_2,x_3,x_4\},{\delta \to x}]
\end{tt}
\end{align*}
As an output we obtain a set of regions $r_i$. Here is
a typical output 
\begin{align*}
\begin{tt}
r=\{\{0, 0, 0, 1\}, \{1, 0, 0, 1\}, \{0, 0, 0, 0\}, \{1, 0, 0, 0\}, \ldots\}
\, .
\end{tt}
\end{align*}
Several comments are in order about the implementation. First, notice that the initial version of the code {\tt asy} was intended for conventional Feynman diagrams
(not a Wilson loop calculation), so that the first two arguments of the routine {\tt WilsonExpand} are reserved for the two basic polynomials in the Feynman parametric 
representation. To apply this routine in the present calculation, it is necessary to reveal all factors of the denominator and then distribute them as factors among the first 
two places. Second, one can put them all in the first place and set 1 in the second place or vice versa. Third, one does not have to provide powers of any polynomials: 
existence of a region does not depend on it! 

Next, according to the prescriptions of the expansion by regions, the contribution of a given region $r_j$ is obtained then from the initial integral by the following
three steps
\begin{align}
\label{MofRrules}
\begin{array}{lll}
& \mbox{(i) rescaling the variables as:} & x_i \to \delta^{(r_j)_i} x_i \, , \quad i=1,\ldots,4 \, , \\[2mm]
& \mbox{(ii) multiplication of the integrand by:} & \delta^{\sum_{i=1}^4 (r_j)_i} \, , \\[2mm]
& \mbox{(iii) expansion of the integrand as:} & \delta\to 0 \, .
\end{array}
\end{align}
As $ \delta\to 0$, each region of parametric integrals obtained in this fashion scales as $\delta^{a+b \eta}$ with some integer $a \geq 0$ and an integer $b$. It is clear 
that we can immediately ignore regions inducing positive $a$'s and take into account only regions with the leading power behavior $\delta^{b \eta}$. The latter produce contributions
of the form $\delta^{b \eta}(C (\eta)+O(\delta))$ with a function $C(\eta)$ that can possess a series 
of poles in $\eta$ as $\eta \to 0$. In our two-loop analysis which 
follows, we encounter $\eta^{j}$ behavior with $j=0,1,-1,-2,-3,-4$. Our goal is to evaluate the leading and subleading logarithms, i.e., $\log^n \delta$ with $n = 4,3,2$. 
Therefore, the problem reduces to the evaluation of the coefficients $C(\eta)$ determined by parametric integrals which are only functions of the auxiliary analytic 
regularization parameter $\eta$. Since we are interested only in double and higher logarithms of $\delta$, it is sufficient to evaluate the coefficients $C_j(\eta)$ in a Laurent 
expansion in $\eta$ and keep only double and higher order poles in $\eta$. Recall that the poles in $\eta$ in the sum of the contributions of all the regions should cancel 
and this is a vital self-consistency check of the entire procedure.

\subsubsection{Example: one loop}

To illustrate the formalism, let us use the integral appearing in the above one-loop example as a case of study,
\begin{align}
\mathcal{J} = 
\int_0^1 ds \int_0^1 d t \int_0^1 d\alpha_1 d \alpha_2 \delta (\alpha_1 + \alpha_2 - 1)
{\rm Int}^{(1)}_{[32],[21]} (x_{12}^{s}, x_{23}^{t}; \alpha_1, \alpha_2)
\, .
\end{align}
It contains all the ingredients relevant to more complex cases without obscuring the method with irrelevant complications.  After the change of variables \re{UnitCubeToRn},
we find
\begin{align}
\mathcal{J} (\delta)
= 
\ft12 (1 - 2 \delta)\int_0^\infty dx_1 \int_0^\infty dx_2 \int_0^\infty dx_3
\frac{x_{3}}{ (1 + x_{3})^2 P(\delta; x_1,x_2,x_3)}
\, ,
\label{example1}
\end{align}
with a rather lengthy polynomial
\begin{align}
P(\delta; x_1,x_2,x_3)&= 
\delta x_{2} + 2 \delta x_{1} x_{2} + \delta x_{1}^2 x_{2} + y x_{3} + \delta x_{1} x_{3} + 
 x_{2} x_{3} + \delta x_{2} x_{3} + x_{1} x_{2} x_{3} + 2 \delta x_{1} x_{2} x_{3} 
\nonumber\\ & 
 + \delta x_{1}^2 x_{2} x_{3} + x_{2}^2 x_{3} + x_{1} x_{2}^2 x_{3} + 
 \delta x_{1} x_{2}^2 x_{3} + \delta x_{1}^2 x_{2}^2 x_{3} + \delta x_{1} x_{3}^2 
\nonumber\\ & 
 + 2 \delta x_{1} x_{2} x_{3}^2 + \delta x_{1} x_{2}^2 x_{3}^2
\, .
\label{example1-P}
\end{align}
Since we are interested only in the leading power behavior, we can replace $(1 - 2 \delta)$ by $1$ in the prefactor from the get-go. 

Next, we define an analytically regularized version of the integral $\mathcal{J} (\delta) \to \mathcal{J} (\delta, \eta)$ as
\begin{align}
\mathcal{J} (\delta, \eta)
=
\ft12 \int_0^\infty dx_1 \int_0^\infty dx_2 \int_0^\infty dx_3
\frac{x_{3}}{ (1 + x_{3})^2 P^{1 - \eta}(\delta; x_1,x_2,x_3)}
\, .
\label{example1-reg}
\end{align}
By running {\tt WilsonExpand} in Mathematica,
\begin{align*}
\begin{tt}
r=WilsonExpand[1+x_3,P, \{x_1,x_2,x_3\},\delta \to x]
\end{tt}
\end{align*}
we obtain eight regions
\begin{align}
\begin{tt}
r = \{\{0, 1, 1\}, \{0, 0, 1\}, \{-1, 1, 0\}, \{-1, 0, 0\}, \{-1, 0, -1\}, \{0,1, 0\}, \{0, 0, 0\}, \{0, 0, -1\}\} \, .
\end{tt}
\end{align}\
Using the prescriptions formulated in \re{MofRrules}, we immediately observe that only four regions $r_j$ with labels $j=3,4,6,7$ are non-zero at leading order, 
with the following scaling behaviors $\{1, \delta^{-\eta}, \delta^{\eta}, 1\}$. Their contribution to $\mathcal{J} (\delta, \eta)$ is given by
\begin{align}
\label{RegJ}
\mathcal{J} (\delta, \eta) = C_3 (\eta) + \delta^{-\eta} C_4(\eta) + \delta^{\eta} C_6 (\eta) + C_7 (\eta) + O(\delta)
\, ,
\end{align}
where the $C$-coefficients are determined by the integral
\begin{align}
C_j (\eta)
=
\ft12 \int_0^\infty dx_1 \int_0^\infty dx_2 \int_0^\infty \frac{dx_3}{(1 + x_{3})^2}
c_j (\eta)
\, ,
\end{align}
with the integrands  
\begin{align}
c_j (\eta)
=
\{
&
x_{1}^{\eta-1} x_{3} (x_{1} x_{2} + x_{3} + x_{2} x_{3} + x_{1} x_{2} x_{3} + x_{3}^2)^{\eta-1}
\, , \\
&
x_{1}^{\eta-1} x_{2}^{\eta-1} x_{3} (x_{1} + x_{3} + x_{1} x_{3} + x_{2} x_{3} + x_{1} x_{2} x_{3})^{\eta-1}
\, ,\nonumber\\
&
x_{3}^{\eta} (1+ x_{1} + x_{1} + x_{1} x_{2} + x_{1} x_{3})^{\eta-1}
\, , 
x_{2}^{\eta - 1} x_3^{\eta} (1+ x_{1})^{\eta - 1}  (1+ x_{2})^{\eta - 1}
\}
\, . \nonumber
\end{align} 
Now we have to evaluate the resulting reduced integrals. This is what we turn to in the next section.

\subsection{Mellin-Barnes technique}

Some of the integrations can be performed analytically for generic values of $\eta$ by the repeated use of a single table integral, i.e., 
\begin{align}
\int_0^\infty dx \, x^\alpha (a + b x)^\beta = 
\frac{\Gamma(1 + \alpha) \Gamma(-1 - \alpha - \beta)}{\Gamma(-\beta)} a^{\alpha + \beta +1} b^{-\alpha-1}\,.
\label{1ltf}
\end{align}
However, a vast majority of nested integrations result in hypergeometric functions. Of course, their poles in $\eta$ can be evaluated by a variety of techniques, but we 
need a method which can be generalized to multifold parametric integrals arising at two loops without the need of advanced case-by-case analyses. 

When evaluating the coefficients $C_j$, we need to resolve singularities in the integration variables $x_i$ in such a manner that the poles in $\eta$ become manifest. For 
parametric integrals stemming from conventional Feynman graphs, there is a plethora of various strategies based on the so-called sector decompositions which are 
implemented in public computer codes to achieve this goal.  However, they cannot be applied (at least, without significant adjustments) to generic parametric integrals, 
like the ones we encounter for the Wilson loop calculation where proper times enter on equal footing with Feynman parameters. In our calculation, we applied for this purpose 
the well-known method of the Mellin-Barnes (MB) representation (see, e.g., Chapter~5 of \cite{Smirnov:2012gma} for a review). The basic tool is summarized by the following simple 
formula
\begin{align}
\frac{1}{[ A_1+A_2 ]^{\nu}} = 
\int_{-i \infty}^{+i \infty}
\frac{dz}{2\pi i}
 \frac{\Gamma(\nu+z) \Gamma(-z)}{\Gamma(\nu)} \frac{A_2^z}{A_1^{\nu+z}} \, ,
\label{MB}
\end{align}
which splits a denominator in terms of preselected components. Here the contour of integration is chosen in a standard way: the poles with a $\Gamma(\ldots+z)$ 
dependence are to the left of the contour and the poles with a $\Gamma(\ldots-z)$ dependence are to its right. 

The above-introduced MB partitioning results in a decomposition of complicated polynomials building up denominators at the cost of introducing extra integrations over the MB 
parameters $z$'s. We apply it iteratively with an aim to obtain a minimal number of resulting MB integrations. Depending on a sequence of partitioning result may vary quite
substantially. The goal is to make the decompositions of the polynomials involved in such a way that the resulting integrals over the $x$-variables could be taken explicitly with the 
help of Eq.\ (\ref{1ltf}).

The end game is that each given coefficient $C_j$ is represented as a multiple MB integral with an integrand containing Euler gamma functions in its numerator and 
denominator. It turns out that such MB integral representations are very convenient to resolve singularities with respect to a given regularization parameter (in particular, 
$\epsilon$ within dimensional regularization or our analytical parameter $\eta$) and obtain results in terms of Laurent series in this parameter. The first step in this procedure 
is a resolution of singularities in the regularization parameter. Roughly speaking, within each of the two known variants of this procedure, one takes residues and shifts integration 
contours to obtain, in the end of the day, a linear combination of multiple MB integrals where the Laurent expansion in the regularization parameter becomes possible directly 
in the {\sl integrand}.

The first strategy of resolving singularities in MB integrals was presented in Ref.\ \cite{Smirnov:1999gc} and the second one in Ref.\ \cite{Tausk:1999vh}. Both of them were 
implemented as computer codes in \cite{Smirnov:2009up} and \cite{Czakon:2005rk}, respectively. Mathematica implemented commands for the resolution of singularities are 
{\tt MBresolve} and {\tt MBcontinue}, correspondingly. In fact, these two versions are compatible in the sense of syntaxis involved and can be used interchangeably. In our 
calculation, our choice was driven by evaluation timing: sometimes one worked faster than the other, although, in most cases, they were essentially compatible.

The above two strategies of resolving singularities in MB integrals are known for more than twenty years and were applied successfully in multiple calculations. We refer to 
Chapter~5 of \cite{Smirnov:2012gma} as well to \cite{HEPforge} for details. This method offers us a real possibility to perform all two-loop calculations in question to the end: 
contributions of regions, expressed in terms of up to six-fold parametric integrals, are reduced to an up to five-fold MB integrals. 
Relying either on {\tt MBresolve} or {\tt MBcontinue} to resolve singularities in $\eta$, we then apply the command {\tt DoAllBarnes} \cite{Kosower:barnesroutines} 
which is based on the first and the second Barnes lemmas and their corollaries. After this, a lot of MB integrations are taken explicitly. 

Most contributions of regions are then evaluated as linear combinations of double and higher poles in $\eta$ with coefficients which are linear combinations of just two elements 
$\{1,\pi^2\}$ accompanied by rational coefficients . Still in some cases, single and even double MB integrations are left. In these situations, we use the possibility to evaluate single and 
double MB integrals numerically with a high precision. For single integrals, we can have the precision of thousand digits and more. For double integrals, the precision of 35 decimal
places is certainly accessible. Since there are only two elements in the basis of numbers involved in our results, we can apply the {\tt PSLQ} algorithm \cite{PSLQ}, for example, 
implemented as a Mathematica built-in command {\tt FindIntegerNullVector}, and present results for the remaining single and double MB integrals in an analytical form. 

\subsubsection{Example: one loop}

In application to the one-loop case, the coefficient $C_3$ can easily be found by means of the 
recursive use of Eq.\ \re{1ltf}. Performing them
in the order $x_{1} \to x_{2} \to x_{3}$, we obtain
\begin{align}
{C}_3(\eta)= & \frac{1}{2} \Gamma(1 - 2 \eta) \Gamma(-\eta) \Gamma(\eta) \Gamma(1 + 2 \eta)
= - \frac{1}{2 \eta^2}+\ldots
\, .
\label{example1-r3-res}
\end{align}
where we left after the second equality sign only the term responsible for double logarithmic asymptotic in the Laurent expansion. Turning next to $C_4$, only one integration can be
done with the help of \re{1ltf}, i.e., with respect to $x_1$, yielding
\begin{align}
{C}_4(\eta)=\frac{\Gamma(1 - 2 \eta) \Gamma(\eta)}{2\Gamma(1 - \eta)} 
\int_0^\infty dx_2 \int_0^\infty \frac{dx_3}{(1 + x_{3})^2 }
\frac{x_{2}^{\eta-1} (1 + x_{2})^{2 \eta-1} x_{3}^{2 \eta}}{(1 + x_{3} + x_{2} x_{3})^{\eta}}
\, .
\label{example1-r4a}
\end{align} 
Next, we proceed with the MB technique. Partitioning the $1 + x_{3} $ and $x_{2} x_{3}$ in its denominator with \re{MB}, we introduce an extra one-fold $z$-integration. However, 
now by changing the order of integrations of $z$ and $x_j$, we can again solve the $x_j$ integrals by the repeated application of Eq.\ (\ref{1ltf}). In this manner, we obtain
\begin{align}
\tilde{C}_4(\eta)=
\int_{-i \infty}^{+i \infty}
\frac{d z}{2 \pi i}
\frac{\Gamma(1 + 2 \eta + z)\Gamma(\eta + z)^2  \Gamma(-z) \Gamma(1 - 3 \eta - z)}{2 \Gamma(2 + \eta + z)}   
\, .
\label{example1-r4b}
\end{align} 
Finally, we apply {\tt MBresolve}/{\tt MBcontinue} to get the pole part in $\eta$
\begin{align}
C_4(\eta)= \frac{1}{2 \eta^2}+\ldots
\, .
\label{example1-r4-res}
\end{align}
Analogous considerations for regions 6 and 7 yield
\begin{align}
C_6 (\eta)= \frac{1}{2 \eta^2}+\ldots
\, , \qquad
C_7 (\eta)= - \frac{1}{2 \eta^2}+\ldots
\end{align}

Summing up all nonvanishing regions together in \re{RegJ} and expanding in $\eta$, we find that all poles cancel, leaving just the finite part
\begin{align}
\mathcal{J} = \frac{1}{2} \log^2 \delta + \dots
\, .
\end{align}
Upon the substitution $\delta \to 1/\Delta$, we recover our result for the diagram in Fig.\ \ref{12loopWLpic} (a)
\begin{align}
\vev{W_4}^{(1)}_{{\rm Fig.} \ref{12loopWLpic} (a)}
=
- 
\frac{N_c}{(4 \pi)^2} ( \ell^2 + O(\ell) )
\, ,
\end{align}
in the multicolor limit, where we introduced
\begin{align}
\ell \equiv \log \Delta \, , \qquad \Delta \equiv \frac{x_{13}^2}{m^2}
\, .
\end{align}
Adding up the other three cusp contributions, we reproduce the well-known results for the off-shell Sudakov behavior \re{OffShellWL1loop}.

\subsection{Webs}

A significant reduction in the number of Feynman graphs one needs to calculate is achieved by means of non-Abelian exponentiation theorems 
\cite{Gatheral:1983cz,Frenkel:1984pz}. Applied to the Wilson loop in question which admits the perturbative series \re{WLPT}, it can be summarized
by the following formula
\begin{align}
\log \vev{W_N} =g^2_{\scriptscriptstyle\rm YM} \vev{W_N}^{(1)} + g^4_{\scriptscriptstyle\rm YM}  \vev{W_N}^{(2)}|_{\rm webs} + \dots \,,
\, .
\end{align}
The one-loop term simply exponentiates, while the two-loop quantum effects contribute only through maximally non-Abelian webs. What this implies is that only
contributions proportional to the color factor $C_F N_c$ have to be selected from corresponding Feynman graphs. For the case at hand, it is rather straightforward
to understand that the Abelian $C_F^2$-structure of $\vev{W_N}^{(2)}$ is already accounted for by the one-loop term since the (Abelian) gluons can be attached
in any order to Wilson links. In other words
\begin{align}
\vev{W_N}^{(2)}
=  
\frac{1}{2} \left[ \vev{W_N}^{(1)} \right]^2
+
\vev{W_N}^{(2)}|_{\rm webs}
\, .
\end{align}
Thus, we can focus only on the second term. Feynman graphs which possess the maximally non-Abelian color structure $C_F N_c$  at two loops, on the one hand, 
and produce at a least double logarithmic dependence on $\Delta$, on the other, are shown in Fig.\ \ref{12loopWLpic} $(b-f)$.

\subsection{Self-energy insertion}

Let us start our two-loop calculation with the graph $(b)$ in Fig.\ \ref{12loopWLpic}, which involves gluon polarization tensor from various fields of the theory and induces
$\mathcal{W}_2^{(2)}$. Since $w_2$ is accompanied by two powers of the YM coupling, we have to expand everything else to order $O (g_{\scriptscriptstyle\rm YM}^2)$, i.e., 
these are terms coming from interactions. There are several sources for gluon couplings. Namely, the gluon self-coupling from the non-Abelian gluon field strength tensors in 
the differential insertion $\Delta^{(2)}$ as well as the three-gluon coupling in the gauge-fixed action $S$, i.e., 
\begin{align}
\label{BWexpansion}
\e^{i S} = \e^{i S_0} \left( 1 - i g_{\scriptscriptstyle\rm YM} \mathcal{V}_{\rm g} + \dots \right)
\, ,
\end{align}
where 
\begin{align}
\mathcal{V}_{\rm g} = \int d^D x_0 V_{\rm g} (x_0) \, , \qquad V_{\rm g} = f^{abc} (\partial_M A_N^a) A_M^b A_N^c
\, .
\end{align}
On the other hand, there is just one source for fermions, scalars and ghosts in the path integral \re{WLVEVinsertions} to this order in coupling: it is the Boltzmann weight.
Their effects appear through the substitution $\mathcal{V}_{\rm g} \to  \mathcal{V}_{\rm g} + \mathcal{V}_{\rm gh} + \mathcal{V}_{\rm f} + \mathcal{V}_{\rm s}$ in Eq.\ 
\re{BWexpansion}. We will not list explicitly other three-field vertices from the $\mathcal{N}=4$ Lagrangian since a reader can figure them out of his/her own properly 
accounting for their normalization relative to the three-gluon one. The notations for those are self-explanatory. Combining everything together, we deduce $\mathcal{W}_2^{(2)}$
\begin{align}
\mathcal{W}_2^{(2)}
=
\frac{1}{2}
\VEV{ w_2
\left[
\frac{1}{2} [\mathcal{V}_{\rm g}^2 + \mathcal{V}_{\rm gh}^2 + \mathcal{V}_{\rm f}^2 + \mathcal{V}_{\rm s}^2] \left( S_0^2  -2 i S_0 \right)
- 2 i \mathcal{V}_{\rm g}^2  S_0
- 3 \mathcal{V}_{\rm g}^2
\right]
}^{(0)}
\, .
\end{align}
Notice the combination $\left( S_0^2  - 2 i S_0 \right)$ accompanying the first term. It has a very simple manifestation in Feynman rules. A simple analysis demonstrates
that the linear term in $S_0$ cancels exactly the quadratic $S_0^2$ when it is inserted into the same gluon line, i.e., 
\begin{align}
S_0^2|_{\rm same \; gluon} = 2 i S_0
\, .
\end{align}
This property is well known, see, e.g., Ref.\ \cite{Arutyunov:2003ad}.
So far all ingredients entering the vacuum expectation value are $D$-dimensional. As a next step, we will integrate out all factors of the free action in the path integral 
making use of the formulas like \re{IntegratingOUTs0}. Namely, we employ
\begin{align}
&
\VEV{ w_2
\mathcal{V}_{\rm g}^2  \left( S_0^2  -2 i S_0 \right)
}^{(0)}
= 12 i^2
\VEV{ w_2 \mathcal{V}_{\rm g}^2 }^{(0)}
\, , \nonumber\\
&
\VEV{ w_2
\mathcal{V}_{\alpha \neq {\rm g}}^2 \left( S_0^2  -2 i S_0 \right)
}^{(0)}
= 2 i^2
\VEV{
\mathcal{V}_{\alpha \neq {\rm g}}^2
}^{(0)}
\, , \\
&
\VEV{ w_2
\mathcal{V}_{\rm g}^2 S_0
}^{(0)}
= 4 i
\VEV{ w_2 \mathcal{V}_{\rm g}^2 }^{(0)}
\, , \nonumber
\end{align}
to obtain  
\begin{align}
\mathcal{W}_2^{(2)}
=
\frac{1}{2} 
\VEV{ w_2
[- \mathcal{V}_{\rm g}^2 - \mathcal{V}_{\rm gh}^2 - \mathcal{V}_{\rm f}^2 - \mathcal{V}_{\rm s}^2]
}^{(0)}
\, .
\end{align}
This is of course the usual form of vacuum polarization from conventional perturbation theory.

Now, we are in a position to dimensionally reduce this expression and introduce the off-shellness. Of course, the latter will regularize short-distance divergences stemming
from integrations in the vicinity of the cusps but it will not regularize ultraviolet singularities stemming from closed internal loops. Since we want to keep the majority of calculations 
in four dimensions we will choose this route but then we will have to introduce a renormalization procedure for internal divergent subgraphs. Another way would be to perform an 
analytic continuation in $\varepsilon$ to negative values $\varepsilon = - \epsilon$ and continue with this extra regulator in addition to $m$. We will comment on it at the end of this 
section.

Adding up individual loops spelled out in the Appendix \ref{PolarizationAppendix}, we find the self-energy insertion into the one-loop Wilson graph
\begin{align}
\label{W22}
\mathcal{W}_{2}^{(2)}
&=
- 
\frac{2 N_c C_F}{(2\pi)^8}
\int d x_1^{\mu_1} d x_2^{\mu_2}
\int d^4 x_0 d^4 x_{0'}  D(x_0; x_1) (-12) \Pi^{\rm tr}_{\mu_1\mu_2} (x_0, x_{0'}) D(x_0; x_2)
\, ,
\end{align}
which is defined in terms of the transverse tensor
\begin{align}
\label{Pitr}
\Pi^{\rm tr}_{\mu_1\mu_2} =
\frac{1}{x_{00'}^6}
\left(
g_{\mu_1 \mu_2} - 2 \frac{(x_{00'})_{\mu_1} (x_{00'})_{\mu_2}}{x_{00'}^2}
\right)
\, .
\end{align}
This expression is gauge invariant and enjoys conformal symmetry as well. Emergence of the latter should not be a surprise since the position space renormalization
affects only coincident points by introducing singular delta functions, while away from them renormalized and bare amplitudes are identical. It is an established fact 
that a two-point function, if conformal invariant, must be gauge invariant too \cite{Schreier:1971um,Freedman:1992tz}.

\subsubsection{Differential renormalization}

Obviously the polarization tensors introduced in the previous section possess short-distance ultraviolet singularities as $x_{0'} \to x_0$ due to products of virtual 
bulk-to-bulk propagators, which are not tempered distributions. So these need to be renormalized. As we already pointed out before, one way would be to use 
$D$-dimensionally regularized version of 
all propagators $D_\epsilon(x,0) = \Gamma (1- \epsilon)/(x^2)^{1 - \epsilon}$ with $D = 4 - 2 \epsilon$ and then subtract $1/\epsilon$-poles in a conventional manner. In fact, 
we would have to resort to dimensional reduction in order to preserve supersymmetry. This procedure sets up a particular renormalization scheme. Actually pole subtraction 
could be done even before any integrations are performed, i.e., at the integrand level. Namely, making use of the Laurent expansion for a product of regularized internal 
propagators, one finds  \cite{Dunne:1992ws}
\begin{align}
\label{DimReProduct}
\mu^{-2\epsilon} D_\epsilon^2(x,0) 
= 
\frac{\pi^2}{\epsilon} \delta^{(4 - 2\epsilon)} (x) - \frac{1}{4} \Box \frac{\log \left[x^2/ (\pi  \mu^2 \e^{- 2 - \gamma_{\rm E}}) \right]}{x^2}
+ O(\epsilon)
\, ,
\end{align}
where $\Box \equiv \partial^2$. One can now subtract poles at the integrand level. We would like to avoid this step altogether since we prefer 
to stay in four dimensions and deal with all graphs, ultraviolet divergent or not, on equal footing.

A renormalization procedure which does not introduce a regulator at an intermediate stage and which is very well tailored to our application
in the position space is the so-called differential renormalization \cite{Freedman:1991tk}. The idea behind this scheme is to rewrite products of propagators 
as derivatives of less  singular functions with derivatives understood via the theory of distributions: formally integrating by parts and moving them to act on
smooth test functions. This step implicitly executes divergent subtractions. The resulting expressions are then identical to the original ones away from singular 
points but have a well-defined behavior when these do sit on top of each other.

Namely, the product of two propagators, which, as we know, does not have a four-dimensional Fourier transform, is replaced with \cite{Freedman:1991tk}, 
cf.\ Eq.\ \re{DimReProduct} above,
\begin{align}
D^2(x,0) 
\to 
D_{\rm R}^2(x,0) 
= 
- \frac{1}{4} \Box \frac{\log \left[ x^2 / \mu^2 \right]}{x^2}
\, .
\end{align}
This expression possesses a Fourier transform when one drops divergent surface terms in the integration by part of the d'Alambertian in the distribution theory sense. 
Higher powers of the inverse squared distances are understood as derivatives of $D_{\rm R}^2(x,0)$. An ambiguity in the renormalization scale $\mu^2\to \e^a \mu^2$ 
results in a local $\sim a \delta^{(4)} (x)$ modification of the above relation and performs a scheme transformation.

Since, in principle, one could introduce different renormalization scales for different contributions to the integrand, there is an a priory intrinsic ambiguity in the differential 
renormalization scheme. However, it can be fixed in gauge theories making use of Ward-Takahashi identities between various graphs. This is what is known as the 
constrained differential renormalization \cite{delAguila:1997kw}. Redoing the calculation that led to Eq.\ \re{Pitr} for $\Pi^{\rm tr}_{\mu_1\mu_2}$, we found according to the 
rules of Ref.\ \cite{delAguila:1997kw}
\begin{align}
\Pi^{\rm \tr}_{\mu_1\mu_2} (x, 0)
=
\left( 
\partial_{\mu_1} \partial_{\mu_2}
-
g_{\mu_1 \mu_2} \Box 
\right) 
\left[
- \frac{5 N_c - 2 n_{\rm f} - n_{\rm s}/2}{6 N_c}
D^2_{\rm R} (x,0)
+ 
\frac{N_c - n_{\rm f} + n_{\rm s}/2}{9 N_c}  \pi^2 \delta^{(4)} (x)
\right]
\, ,
\end{align}
where we kept contributions from gluons/ghosts, $n_{\rm f}$ fundamental fermions and $n_{\rm s}$ scalars separately. This result agrees with
corresponding expressions in Refs.\ \cite{Perez-Victoria:1998gla} and \cite{delAguila:1998nd} for the gluon-fermion and scalar terms, respectively.
The reason why we kept effects of different particle types apart is to notice that in supersymmetric YM theories with $\mathcal{N} =1,2,4$ supercharges,
the contact term vanishes identically. It is an interesting fact akin to the vanishing of $2\epsilon$-dimensional contributions to the tensor structure
of the polarization tensor in the dimensional reduction scheme \cite{Belitsky:2003ys}.

Making use of the differential renormalization, we find for \re{W22}
\begin{align}
\label{W22renormalized}
\mathcal{W}_{2}^{(2)}
=
- \frac{4 N_c C_F}{(4 \pi)^4} \log(\mu^2/m^2) \left[ \ell^2  + O (\ell) \right]
\, ,
\end{align}
where $\mu$ is an arbitrary renormalization scale. So this contribution requires a scale setting procedure! For physical processes, there are several prescriptions
to handle it, see, e.g., Refs.\  \cite{Grunberg:1980ja,Stevenson:1980du,Brodsky:1982gc}. The Wilson loop in question is not per se an observable, so it enjoys even 
more freedom. Let us motivate our choice. In the well-studies massless case, one chooses
to regulate the cusp and UV divergences in the same fashion by means of dimensional regularization. But even in that case there is an ambiguity in
the scale setting which is typically solved in a minimal fashion by identifying $\mu_{\rm UV} = \mu_{\rm cusp}$. But this is in no way unique since one
can introduce a finite renormalization by setting $\mu_{\rm UV} = {\rm e}^a \mu_{\rm cusp}$ with some constant $a$ (recall the MS and DR subtractions
in dimensional reduction regularization \cite{Belitsky:2003ys}). In the current ``massive'' case, the situation is somewhat similar. Namely, the short-distance cusp 
divergences are regularized by the off-shellness $m^2 = x_{j,j+1}^2$. So what should one choose for $\mu^2$? If one sets it equal to ${\rm e}^a m^2$, the 
self-energies will possess at most $\ell^2$-behavior, but it will be a pure finite renormalization with no higher logarithms to induce it. It is known that in dimensional regularization, 
one generates an extra overall logarithm from divergent subgraphs after subtraction on top of the ones of the parent graph, see, e.g., \cite{Drummond:2007cf}. For the 
off-shell case, the reductionism suggests that since $\mu^2$ should be non-vanishing even in the massless case, yet another option is to set it to $x_{13}^2 = x_{24}^2$. 
Then the self-energy graph will contain a triple logarithm as expected. More generally, we can use the freedom in the scale setting procedure of ultraviolet divergences 
to have its mass dimension carried by the off-shellness but accompanied by a power of the only scaling variable in the problem, i.e., $\Delta = x_{13}^2/x_{12}^2$. 
Therefore, we introduce the following parametrization
\begin{align}
\mu^2 = {\rm e}^{a} m^2 \Delta^{- \alpha}
\, ,
\end{align}
with some rationals\footnote{Scheme transformation constants are of lower degree of transcendentality than a given loop order can produce.} $a$ and $\alpha$, so that 
\begin{align}
\mathcal{W}_{2}^{(2)}
=
\frac{4 N_c C_F}{(4 \pi)^4} \left[ \alpha \ell^3  - a \ell^2 + O (\ell) \right]
\, .
\end{align}
So we kill two birds with one stone: we introduce a scale setting and a finite renormalization with a choice of these two numbers. These will be fixed below.

Before we close this section, let us mention that we performed the above calculation in the dimensional reduction scheme as well making use of the momentum-space
polarization operator of Ref.\ \cite{Belitsky:2003ys} and Fourier transforming it to the position space. After subtraction of the ultraviolet pole in $\varepsilon$, we ended up with the
subtracted form of Eq.\ \re{W22} which differs from \re{W22renormalized} by a rescaling of $\mu^2$, as expected.

\subsection{Abelian cross ladders }

Let us turn next to the diagrams $(c,d)$ of Fig.\ \ref{12loopWLpic} with the Abelian cross ladders. Since all powers of the YM coupling are already accounted for by the Wilson
links, the differential operator insertion $\Delta^{(2)}$ has to be taken at tree level.  The generic form of this contribution to $\mathcal{W}_4^{(2)}$ is
\begin{align}
\mathcal{W}_4^{(2)}
=
- 
\frac{1}{2}
\VEV{ w_4 (S_0^2 - 2 i S_0) }^{(0)}
\, .
\end{align}
For the diagram $(c)$, we find immediately its contribution to the web
\begin{align}
\left. \mathcal{W}_4^{(2)} \right|_{{\rm Fig.} \ref{12loopWLpic} (c)}
=
&-
\frac{N_c C_F}{2 (2 \pi)^4} \int_0^1 ds_2 \int^{s_2}_0 ds_1  \int_0^1 dt_2 \int^{t_2}_0 dt_1 \int_0^1 d\alpha_1 d \alpha_2 \delta (\alpha_1 + \alpha_2 - 1)
\\
&
\times
\int_0^1 d\beta_1 d \beta_2 \delta (\beta_1 + \beta_2 - 1)
{\rm Int}^{(1)}_{[32],[21]} (x_{12}^{s_1}, x_{23}^{t_1}; \alpha_1, \alpha_2)
{\rm Int}^{(1)}_{[32],[21]} (x_{12}^{s_2}, x_{23}^{t_2}; \beta_1, \beta_2)
\, , \nonumber
\end{align}
where only the $N_c C_F$ color factor is kept. Here the integrand given by the product of two one-loop integrands \re{1loopIntegrand}. Similarly, for the graph $(d)$, we get
\begin{align}
\left. \mathcal{W}_4^{(2)} \right|_{{\rm Fig.} \ref{12loopWLpic} (d)}
=
&-
\frac{N_c C_F}{2 (2 \pi)^4} \int_0^1 d s \int_0^1 dt_2 \int^{t_2}_0 dt_1  \int^{1}_0 d r \int_0^1 d\alpha_1 d \alpha_2 \delta (\alpha_1 + \alpha_2 - 1)
\\
&
\times
\int_0^1 d\beta_1 d \beta_2 \delta (\beta_1 + \beta_2 - 1)
{\rm Int}^{(1)}_{[32],[21]} (x_{12}^{t_1}, x_{23}^{r}; \alpha_1, \alpha_2)
{\rm Int}^{(1)}_{[21],[14]} (x_{41}^{s}, x_{12}^{t_2}; \beta_1, \beta_2) 
\, . \nonumber
\end{align}
The latter integral enjoys both hard scales $x_{13}^2$ and $x_{24}^2$, but as we advocated earlier all we need is its value at the symmetric point $x_{13}^2 = x_{24}^2$.
Evaluation of these contributions yields
\begin{align}
\left. \mathcal{W}_4^{(2)} \right|_{{\rm Fig.} \ref{12loopWLpic} (c)}
&
=
\frac{N_c C_F}{(4 \pi)^4} \left[ - \ft{1}{2} \ell^4 + O (\ell)\right]
\, , \\
\left. \mathcal{W}_4^{(2)} \right|_{{\rm Fig.} \ref{12loopWLpic} (d)}
&
=
\frac{N_c C_F}{(4 \pi)^4}  \left[ -2 \zeta_2 \ell^2 + O (\ell) \right]
\, .
\end{align}
Notice that the integral in Fig.\ \ref{12loopWLpic} $(c)$ does not have cubic or quadratic terms following the quartic: these cancel between various contribution in the product of 
the two terms of Eq.\ \re{1loopIntegrand}. Though, the second terms in ${\rm Int}$'s are proportional to $m^2$, nevertheless they induces logarithms of lower degree when
multiplied by the first ones.

\subsection{Three-gluon vertex}

Finally, the three-gluon vertex contribution to the Wilson loop average reads
\begin{align}
\label{3gluonW}
\mathcal{W}_3^{(2)} = \frac{1}{2}\VEV{w_3 \mathcal{V}_3 \left( - 2 i + 4 S_0 + i S_0^2 \right)}^{(0)}
\, .
\end{align}
Since we have an inner interaction vertex, we can integrate out the free YM action in the path integral. As can be anticipated the last two terms in round braces cancel against 
each other via the identity
\begin{align}
\VEV{w_3 \mathcal{V}_3 S_0^2 }^{(0)} = 4 i \VEV{w_3 \mathcal{V}_3 S_0 }^{(0)} 
\, ,
\end{align}
leaving only the first term of the regular perturbative series
\begin{align}
\mathcal{W}_3^{(2)} = \VEV{w_3 (- i) \mathcal{V}_3}^{(0)}
\, .
\end{align}

Now that we exhausted the use of the equations of motion, we can dimensionally reduce this expression and get the following parametric integral for the diagram in 
Fig.\ \ref{12loopWLpic} $(e)$
\begin{align}
\mathcal{W}^{(2)}_{3}|_{{\rm Fig.} \ref{12loopWLpic} (e)}
=
\frac{N_c C_F}{16 (2 \pi)^2} 
\int_0^1 ds_2 \int^{s_2}_0 ds_1  \int_0^1 dt  \int_0^1 d\alpha_1 d \alpha_2 d \alpha_3 \delta (\alpha_1 + \alpha_2 + \alpha_3 - 1) \frac{n_{\rm (e)}}{L^2_{[32];[21]^2}}
\, ,
\end{align}
where the numerators is
\begin{align}
n_{\rm (e)} =  t ( \alpha_2 - \alpha_1) \alpha_3 x_{13}^2 (x_{13}^2 - 4 m^2)
\, ,
\end{align}
while the denominator reads
\begin{align}
L_{[32];[21]^2}
&
= \alpha_1 \alpha_2 m^2 (s_1-s_2)^2
+
\alpha_1 \alpha_3 \left( \bar{s}_1 t x_{13}^2 +m^2 (t - \bar{s}_1)^2 \right)
\\
&
+
\alpha_2 \alpha_3 \left( \bar{s}_2 t x_{13}^2 + m^2 (t-\bar{s}_2)^2  \right)
+
m^2 \left( \alpha_1 s_1 \bar{s}_1+  \alpha_2 s_2 \bar{s}_2+  \alpha_3 t \bar{t} \right)
\, . \nonumber
\end{align}
Their calculation according to the procedure outlined above yields
\begin{align}
\mathcal{W}^{(2)}_{3}|_{{\rm Fig.} \ref{12loopWLpic} (e)}
=
\frac{N_c C_F}{(4 \pi)^2} 
\left[ 
\ft{1}{4} \ell^4 - \ft{3}{2} \ell^3 + \left( 1 + \ft32 \zeta_2 \right) \ell^2 + O (\ell)
\right]
\, .
\end{align}

The last graph to address is Fig.\ \ref{12loopWLpic} $(f)$. Since the gluons dress up two cusps rather than one, the resulting integral
\begin{align}
\mathcal{W}^{(2)}_{3}|_{{\rm Fig.} \ref{12loopWLpic} (f)}
=
\frac{N_c C_F}{16 (2 \pi)^2} 
\int_0^1 ds  \int_0^1 dt   \int_0^1 dr \int_0^1 d\alpha_1 d \alpha_2 d \alpha_3 \delta (\alpha_1 + \alpha_2 + \alpha_3 - 1) \frac{n_{\rm (f)}}{L^2_{[43];[32];[21]}}
\, ,
\end{align}
is function of the two hard scales $x_{13}^2$ and $x_{24}^2$ entering in the numerator
\begin{align}
n_{\rm (f)} 
&
=  \alpha_3  x_{13}^2\big( (\alpha_1 + \bar{t}\alpha_2) x_{24}^2  + \bar{s} \alpha_1 (x_{13}^2 + x_{24}^2 - 4 m^2) \big)
\nonumber\\
&
+ \alpha_1  x_{24}^2 \big( (\alpha_3 + t \alpha_2) x_{13}^2  + r \alpha_3 (x_{13}^2 + x_{24}^2 - 4 m^2) \big)
\nonumber\\
&\qquad
-  \alpha_2 (x_{13}^2 + x_{24}^2 - 4 m^2)  (\bar{s} \alpha_1 x_{13}^2 + r \alpha_3 x_{24}^2)
\, ,
\end{align}
as well as the denominator
\begin{align}
L_{[43];[32];[21]}
&
=
\alpha_1 \alpha_2  \left( \bar{s} t x_{13}^2 +m^2 (\bar{s} - t)^2 \right)
+
\alpha_2 \alpha_3 \left( \bar{t} r x_{24}^2 +m^2 (\bar{t} - r)^2 \right)
\\
&
+
\alpha_1 \alpha_3 \left( \bar{s} \bar{r} x_{13}^2 +s r x_{24}^2 +m^2 (s - r)^2 \right)
+
m^2 \left( \alpha_1 s \bar{s}+  \alpha_2 t \bar{t}+  \alpha_3 r \bar{r} \right)
\, . \nonumber
\end{align}
The three-term numerator is a reflection of the the three tensor structures of the three-gluon vertex. Analyzing the small-$m$ asymptotic behavior of this
graph at the symmetric point $x_{13}^2 = x_{24}^2$, we find that the last term in the numerator is $O(\ell)$, while the other two give identical contributions.
Summing these up, we deduce
\begin{align}
\mathcal{W}^{(2)}_{3}|_{{\rm Fig.} \ref{12loopWLpic} (f)}
=
\frac{N_c C_F}{(4 \pi)^2} 
\left[  \zeta_2 \ell^2 + O (\ell) \right]
\, .
\end{align}

\subsection{Sudakov scaling}

Adding all graphs up together accompanied by the corresponding multiplicity factors, we observe an immediate cancellation of the $\ell^4$ dependence. To cancel $\ell^3$,
we have to properly choose the coefficient $\alpha$ in the scale setting procedure for $\mu^2$: this is achieved for $\alpha = 3/4$. Finally, the rational contribution
to the coefficient of $\ell^2$ is scheme dependent and can be eliminated by equating $a = 1/2$. The transcendental portion of this coefficient is scheme independent
and is the main result of this section. Combined with the one-loop asymptotics, the Sudakov behavior is driven by the exponent
\begin{align}
\log\vev{W_4} =  - (4 g^2 - 4 \zeta_2 g^4) \ell^2 + O (\ell) 
\, ,
\end{align}
in the planar limit. Surprisingly enough, we find that the function of the coupling accompanying $\ell^2$ is $\Gamma_{\rm hex}$ shown in Eq.\ \re{ADs}, which made its first 
appearance in the hexagon remainder function when all cross ratios tend to zero \cite{Basso:2020xts}. Of course, to this order in coupling we cannot clearly differentiate 
whether it is truly $\Gamma_{\rm hex}$ or rather a linear combination of the cusp and hexagon anomalous dimensions (or of all three) that governs the asymptotic behavior
\footnote{We are grateful to Benjamin Basso and Lance Dixon for a discussion of this issue.}. A three-loop calculation would resolve this predicament. However, it is currently 
out of reach since it requires a three-loop massive calculation.

\section{Conclusions}

In this paper, we proposed a generalization of the bosonic Wilson loop on a piece-wise contour to the case of the cusps separated in Minkowski space by
timelike, rather than null, intervals. The motivation behind it was to find a generalization of the scattering amplitudes/Wilson loop duality, well-known for on-shell 
massless setup, to the off-shell situation. This was achieved by considering a holonomy of the gauge connection just above four dimensions and subsequent 
dimensional reduction of its interior down to four with the extra dimensions of the perimeter accommodating the off-shellness. Practically this was done by means 
of the formalism of Lagrangian insertions which clearly disentangle the boundary from the bulk.

Performing a one-loop calculation of the off-shell Wilson loop for four sites we demonstrated its exact equivalence to the four-gluon amplitude in the Coulomb branch 
of $\mathcal{N} = 4$ SYM \cite{Caron-Huot:2021usw}. This implies that the former enjoys conformal symmetry to this order in coupling. By calculating its asymptotics 
of vanishing virtuality we uncovered a well-known form of the Sudakov behavior with a factor of 2 difference between the on- and off-shell amplitudes at this order. 
This repeated and confirmed the familiar Abelian Sudakov form factor story known for almost half a century. However, it appears that it was taken for granted that this 
doubling is the only effect that occurs at higher loop orders for the infrared asymptotics of scattering amplitudes for dimensional versus off-shell regulator, see 
\cite{Korchemsky:1988hd,Drummond:2007aua}. Results of Ref. \cite{Caron-Huot:2021usw} suggested instead a different mechanism and pointed to a completely 
different function of the coupling that drives the infrared evolution of amplitudes in the off-shell regime, and this is the octagon anomalous dimension  $\Gamma_{\rm oct}$, 
see Eq.\ \re{ADs}. 

Due to availability of perturbative data for the Coulomb branch scattering amplitudes through octagons alluded to above, we calculated the off-shell Wilson loop at $O(g^4)$ 
in 't Hooft coupling. We considered only its Sudakov limit at a symmetric point which is sufficient to deduce the anomalous dimension which governs it. The analysis was
involved and subtle. The non-vanishing off-shellness regularizes only short-distance divergences around the perimeter of the loop, i.e., cusp singularities. However, the interior 
quantum loops had to be dealt with separately: they had to be renormalized by an independent procedure. We chose differential renormalization as such since it leaves the 
dimensionality of the interior intact. Subtraction of emerging ultraviolet divergences lead to the introduction of yet another independent mass scale in addition to the off-shellness.
We observed that upon a proper choice of the renormalization scale setting procedure, the off-shell Wilson loop enjoys exponentiation of the Sudakov double logarithms,
however, the accompanying coefficient was found to be $\Gamma_{\rm hex}$, see Eq.\ \re{ADs}. The latter was first discovered in the origin limit of the remainder
function of on-shell six gluon amplitude in Ref.\ \cite{Basso:2020xts}. Does it imply that the conformal symmetry is violated starting from two loops? Of course, all that conformal
symmetry says about the structure of the Wilson loop is that it is a function of two conformal cross ratios, while the dependence on the 't Hooft coupling is completely arbitrary
from this standpoint. To address this question in a systematic fashion, we have to restore generic kinematics in order to elucidate the role of perimeter regularization in the 
pattern of its breaking, if any. It can also be studied by means of conformal Ward identities along the lines of Refs.\ \cite{Belitsky:1998gc,DruHenKorSok07}. Another natural 
question which arises is whether the exponent governing the current Sudakov behavior is universal. Is it the same for the off-shell Wilson loop with any number of sites? This 
question can be addressed in a rather straightforward manner making use of the technique employed in this paper.

A posteriori, it appears dubious for the octagon anomalous dimension to even emerge from a bosonic Wilson loop calculation, especially at higher loops. The perturbative expansion 
of $\Gamma_{\rm oct}$ contains only even zeta values, while Wilson loops pour-forth both, even and odd. So were the octagon anomalous dimension to stem from one, 
what constraints on its form and perturbative series would ensure this? 

While we apparently have a robust definition of the off-shell Wilson loop order-by-order in the perturbative expansion, can it be lifted nonperturbatively? If it does,
it would be interesting to to understand the dynamics of the emerging two-dimensional world-sheet. The energy-density of the gauge flux sourced by the contour is different 
from either the cusp or the octagon anomalous dimensions. It is known that all three $\Gamma_\alpha$'s are accommodated in the same equation with a tilted flux-tube 
kernel \cite{Basso:2020xts}. Is there a tilted form of a long-range Baxter equation which is their progenitor and what kind of integrable model it corresponds to? If this problem 
has an affirmative solution, it would be interesting to explore the spectrum of excitation propagating on the tilted world-sheet, their dispersion relations and scattering matrices.
Hopefully, some of these questions can be answered in the not too far distant future.

\section*{Acknowledgments}

We would like to thank Benjamin Basso, Simon Caron-Huot, Frank Coronado, Lance Dixon and Yorgos Papathanasiou for instructive correspondence and valuable comments on the 
manuscript. We are grateful to Alexander Smirnov for his help with codes used in this work. The work of A.B.\ was supported by the U.S.\ National Science Foundation under the grant 
No.\ PHY-2207138.
The work of V.S.\ was supported by the Ministry of Education and Science of
the Russian Federation as part of the program of the Moscow Center for Fundamental and
Applied Mathematics under Agreement No.\ 075-15-2019-1621.

\appendix

\section{Mellin-Barnes representation}

Let us provide details for the reduction of parametric integrals in the body of the paper to contour integral representations. We will focus on the most complicated exchange contribution
in Fig.\ \ref{OneLoopPic} $(a)$ and merely state results for the rest. We split it according to the structure of the numerator in terms of its $z^2$-dependence, i.e., coefficients accompanying 
$z_{13}^2 z_{24}^2$, $z_{13}^2$, $z_{24}^2$ and the rest. The starting point is the well-known formula \cite{Belokurov:1983rkp,Davydychev:1990jt}
\begin{align}
\frac{1}{[A_1 + A_2 + A_2]^\nu} = \frac{1}{A_3^\nu} \int \frac{d z_1 d z_2}{(2 \pi i)^2} \left( \frac{A_1}{A_3} \right)^{z_1} \left( \frac{A_2}{A_3} \right)^{z_2}
\frac{\Gamma (-z_1) \Gamma (-z_2) \Gamma (z_1 + z_2 + \nu)}{\Gamma (\nu)}
\, ,
\end{align}
(which is a natural generalization of Eq.\ \re{MB} quoted in the main text) with implicit integration contours running along the imaginary axis and separating poles of the Euler 
Gamma functions with positive and negative signs of $j$'s. We apply it to the denominator $\mathcal{D}$ defined in Eq.\ \re{ExchangeDenominator}. After this, the integral over 
$\sigma$ can be performed in terms of the Appell function $F_1$ \cite{Srivastava:1995} (see Eq. (3.6) there), which reads
\begin{align*}
\label{AppellF1fromSigma}
\int_{\tau_2}^{\tau_1} d \sigma (\tau_1 - \sigma)^{\alpha - 1} (\sigma - \tau_2)^{\beta - 1}  \sigma^\gamma \bar\sigma^\delta
=
\tau_2^\gamma \bar\tau_2^\delta \tau_{12}^{\alpha + \beta - 1} \frac{\Gamma (\alpha) \Gamma (\beta)}{\Gamma (\alpha + \beta)}
F_1
\left(
\beta, - \gamma, - \delta, \alpha + \beta, - \frac{\tau_{12}}{\tau_2}, \frac{\tau_{12}}{\bar\tau_2}
\right)
\, .
\end{align*}
Unfortunately, we were not able to perform the remaining integrals over the proper times $\tau_1$ and $\tau_2$ analytically due to the coupled nature of the two arguments of the
Appell function. To be able to advance and do it, however, we will proceed with a further Mellin-Barnes representation for the Appell function itself.

Before we do it, however, we observe that for the $z_{13}^2 z_{24}^2$-structure, the Appell function actually reduces to the simpler hypergeometric function ${_2F_1}$ by means of 
the know identity
\begin{align*}
F_1
\left(
\alpha; \beta_1, \beta_2; \beta_1 + \beta_2; \tau_1, \tau_2
\right)
=
(1 - \tau_2)^{-\alpha}
{_2F_1}
\left(
\alpha, \beta_1, \beta_1 + \beta_2, \frac{\tau_{12}}{\bar\tau_2}
\right)
\, ,
\end{align*}
such that
\begin{align}
\left\langle\!\!\!\left\langle
\frac{\Gamma (D/2)}{\mathcal{D}_{x_{13},  x_{24}}^{D/2}}
\right\rangle\!\!\!\right\rangle
&
=
(-m^2)^{-D/2} 
\int_0^1 d \tau_1 \int_0^1 d \tau_2
\int \prod_{j = 1}^2 \frac{d z_j}{2 \pi i} \Gamma (- z_j)
\left( \frac{x_{13}^2}{m^2} \right)^{z_1}
\left( \frac{x_{24}^2}{m^2} \right)^{z_2}
\frac{\Gamma\left(\frac{D}{2}+z_1+z_2\right)^3}{\Gamma (D+2 z_1+2 z_2)}
\nonumber\\
&\times
\tau_1^{z_2}  \bar\tau_1^{- \frac{D}{2} - z_2}   \tau_2^{-\frac{D}{2}-z_1} \bar\tau_2^{z_1}  \, 
{_2F_1} 
\left(
\frac{D}{2}+z_1+z_2,\frac{D}{2}+z_1+z_2; D + 2 z_1+ 2 z_2; - \frac{\tau_{12}}{\bar\tau_1 \tau_2}\right)
\, .
\end{align}
Finally, employing the Mellin-Barnes representation for ${_2F_1}$,
\begin{align}
{_2F_1}
\left(
\alpha, \beta ; \gamma;
\tau
\right)
&
=
\frac{\Gamma (\gamma)}{\Gamma (\alpha) \Gamma (\beta) \Gamma (\gamma-\alpha) \Gamma (\gamma-\beta)}
\nonumber\\
&
\times
\int \frac{d z}{2 \pi i}
(1 - \tau)^z \Gamma (-z) \Gamma (\alpha + z) \Gamma (\beta + z) \Gamma (\gamma - \alpha - \beta - z)
\, ,
\end{align}
we obtain a three-fold contour integral which is quoted in Eq.\ \re{MBtripple} of the main text.

For $z_{13}^2$-coefficient, the $F_1$ does not simplify and is left as it is. However, to employ a well-known contour integral form for it \cite{Appell:1926}, namely,
\begin{align}
F_1 (\alpha; \beta_1, \beta_2; \gamma; \tau_1, \tau_2)
&
=
\frac{\Gamma (\gamma)}{\Gamma (\alpha) \Gamma (\beta_1) \Gamma (\beta_2)}
\nonumber\\
&
\times
\int \frac{dz_1 dz_2}{(2 \pi i)^2}
(-\tau_1)^{z_1} (-\tau_2)^{z_2}
\frac{\Gamma (- z_1) \Gamma (- z_2) \Gamma (\beta_1 + z_1) \Gamma (\beta_2 + z_2) \Gamma (\alpha + z_1 + z_2)}{\Gamma (\gamma + z_1 + z_2)}
\end{align}
one has to make sure that both arguments of the function have phases obeying the constraints $|{\rm arg} (- \tau_{1,2})| < \pi$. This is not the case in Eq.\ \re{AppellF1fromSigma} 
since both integration regions $\tau_1 \lessgtr \tau_2$ contribute on equal footing. The problem can be partially alleviated by reducing the region $\tau_1 < \tau_2$
to the other one $\tau_1 > \tau_2$. In fact the result for the latter just doubles. However, after this is done, while the first argument in Eq.\ \re{AppellF1fromSigma}
satisfies the above condition, the second one does not. To correct this, we use the transformation 
\begin{align*}
F_1
\left(\alpha; \beta_1, \beta_2; \gamma; \tau_1, \tau_2 \right)
=
\left(1-\tau_2\right)^{- \beta_2} \left(1-\tau_1\right)^{-\alpha - \beta_1 + \gamma}
F_1
\left(\gamma - \alpha; \gamma - \beta_1 - \beta_2, \beta_2; \gamma ; \tau_1, \frac{\tau_{12}}{\bar{\tau}_2}\right)
\, .
\end{align*}
Assembling everything together, we obtain the following four-fold integral
\begin{align*}
\left\langle\!\!\!\left\langle 
[x_{24}^2 \sigma + 2 m^2 \bar\sigma]
\frac{\Gamma (D/2)}{\mathcal{D}_{x_{13},  x_{24}}^{D/2}}
-
\frac{2}{\tau_{12}^2} 
\frac{\Gamma (D/2-1)}{\mathcal{D}_{x_{13},  x_{24}}^{D/2-1}}
\right\rangle\!\!\!\right\rangle
=
\frac{4}{(- m^2)^{D/2}}
\int \prod_{j = 1}^4 \frac{d z_j}{2 \pi i} \Gamma (- z_j)
\left( \frac{x_{13}^2}{m^2} \right)^{z_1}
\left( \frac{x_{24}^2}{m^2} \right)^{z_2}
\nonumber
\end{align*}
\vspace{-16pt}
\footnotesize
\begin{align*}
&
\times
\Gamma (z_3+1) \Gamma (z_3+z_4+1) 
\Gamma \left(D/2+z_1+z_2\right) 
\Gamma \left(D/2+z_1+z_2+z_4\right)
\Gamma \left(D/2+z_1+z_2+z_3+z_4\right)
\nonumber\\[5pt]
&
\times
\frac{\Gamma \left(D/2-z_2-z_4+2\right)
\Gamma \left(-D/2+z_1-z_2+z_3+3\right)  \Gamma   \left(-D/2-z_1-z_3-z_4+1\right)}{
\Gamma \left(-D/2-z_2+z_3+3\right) \Gamma (-D-z_2-z_4+4)  \Gamma (D+2 z_1+2z_2+z_3+z_4)}
\bigg[
\frac{2 m^2 (z_3 + 1)}{D + 2 z_1 + 2 z_2 + 2 z_4 - 2}
f_{111}
\nonumber\\
&\qquad\qquad\qquad\qquad\qquad\quad
+
\frac{m^2 (D + 2 z_1+ 2 z_2 - 2)}{D + 2 z_1 + 2z_2 + 2 z_4 -2}
f_{110}
+
\frac{x_{24}^2 (D + 2 z_2 - 2 z_3 - 4) (D + z_2 + z_4 - 3)}{(D + 2 z_2 + 2 z_4 - 2) (D - 2 z_1 + 2 z_2 - 2 z_3 - 4)}
f_{001}
\bigg]
\nonumber
\end{align*}
\vspace{-18pt}
\normalsize
\begin{align}
\label{MBfourfold}
\end{align}
where
\begin{align}
f_{\alpha_1 \alpha_2 \beta} = {_3F_2}
\left.
\left(
{ - \beta - z_2 , \alpha_2 - D/2  + 2 + z_1 - z_2 + z_3 , \alpha_1 - D/2 + 1 - z_2 - z_4 \atop \alpha_2 - D/2  + 2 - z_2 + z_3, \alpha_2 - D + 3 - z_2 - z_4}
\right|
1
\right)
\, .
\end{align}

Finally, let us address the last term in the numerator of Eq.\ \re{EqExchange}, which we referred to in the main body as to the self-energy-like contribution. Indeed in
the exchange graph \ref{OneLoopPic} (a) it can be integrated over the proper times to get
\begin{align}
W_{4(1,a)}^{\rm off-shell} 
&\to \frac{i \Gamma^2 (D/2)}{32 \pi^D} \int_0^1 d \tau_1 \int_0^1 d \tau_2 \int d^D x_0 
\frac{
[x_{01}^2 - x_{02}^2] [x_{03}^2 - x_{04}^2]
}{
[- \bar\tau_1 x_{01}^2 - \tau_1 x_{02}^2]^{D/2}
[- \bar\tau_2 x_{03}^2 - \tau_2 x_{04}^2]^{D/2}
}
\nonumber\\
&=
\frac{i \Gamma^2 (D/2-1)}{32 \pi^D} \int d^D x_0 
\Big( [- x_{01}^2]^{1- D/2} - [- x_{02}^2]^{1 - D/2} \Big)
\Big( [- x_{03}^2]^{1- D/2} - [- x_{04}^2]^{1 - D/2} \Big)
\, ,
\end{align}
and depends only on the boundary values of the links. The last contribution to the graph $(b)$ as well the diagram $(c)$ admit the same form and add up to zero.

Last but not least, the vertex graph is deduced from the exchange by taking the residue at $z_2 = 0$ and setting $x_{24}^2 = 0$. In this case, the hypergeometric function
${_3F_2}$ reduces to products of Euler Gammas and the result reads
\begin{align}
\label{VertexResult}
\left\langle\!\!\!\left\langle
2 z_{13}^2 m^2 \bar\sigma
\frac{\Gamma (\frac{D}{2})}{\mathcal{D}_{x_{13},  0}^{D/2}}
-
\, 2 \frac{z_{13}^2}{\tau_{12}^2}
\frac{\Gamma (\frac{D}{2}-1)}{\mathcal{D}_{x_{13},  0}^{D/2-1}}
\right\rangle\!\!\!\right\rangle
=
\frac{4}{(- m^2)^{D/2-1}}
\int \prod_{j = 1}^3 \frac{d z_j}{2 \pi i} \Gamma (- z_j)
\left( \frac{x_{13}^2}{m^2} \right)^{z_1}
\hspace{45pt}
\end{align}
\vspace{-16pt}
\footnotesize
\begin{align*}
&
\times
\left( 
\frac{\left(\frac{d}{2} - z_1 - z_2 - 3\right) (d+2 z_3-4)}{2 \left(\frac{d}{2}-z_2-3\right) (d+z_3-4)} - \frac{\frac{d}{2}+z_1-1}{z_2+1}-1
\right)
\Gamma (z_2+2) \Gamma (z_2+z_3+1) \Gamma \left(\frac{d}{2}+z_1\right) \Gamma \left(-\frac{d}{2}-z_3+2\right) 
\\
&\times
\frac{\Gamma \left(-\frac{d}{2}+z_1+z_2+3\right) \Gamma
   \left(\frac{d}{2}+z_1+z_3-1\right) \Gamma \left(-\frac{d}{2}-z_1-z_2-z_3+1\right) \Gamma
   \left(\frac{d}{2}+z_1+z_2+z_3\right)}{\Gamma \left(-\frac{d}{2}+z_2+3\right) \Gamma (-d-z_3+4) \Gamma (d+2
   z_1+z_2+z_3)}
\, .
\end{align*}
\normalsize
It was also rechecked by an explicit calculation as well.

\section{Vacuum polarization tensors}
\label{PolarizationAppendix}

\begin{figure}[t]
\begin{center}
\mbox{
\begin{picture}(0,60)(200,0)
\put(0,0){\insertfig{14}{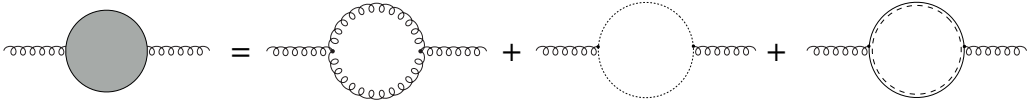}}
\end{picture}
}
\end{center}
\caption{\label{SelfEnergyFig} One-loop vacuum polarization from gluons, ghosts and fermions/scalars.}
\end{figure}

Let us quote here the form of vacuum contributions from gluon, ghost, fermion/scalar fields to the gluon propagator. Calculating corresponding Green functions,
we find 
\begin{align}
\vev{A_{\mu_1}^{a_1} (x_1) A_{\mu_2}^{a_2} (x_2) \mathcal{V}_{\alpha}^2 }^{(0)}
=
- \frac{4 N_c}{(4 \pi^2)^4} \delta^{a_1 a_2}
\int d^4 x_0 d^2 x_{0'}  D(x_1; x_0) \Pi^{\alpha}_{\mu_1\mu_2} (x_0, x_{0'}) D(x_2; x_{0'})
\, ,
\end{align}
where the polarization tensors (in the Feynman gauge)
\begin{align}
\Pi^{\rm g}_{\mu_1\mu_2} (x_0, x_{0'})
&= 10 \frac{g_{\mu_1 \mu_2}}{x_{00'}^6} - 22 \frac{(x_{00'})_{\mu_1} (x_{00'})_{\mu_2}}{x_{00'}^8}
\, , \\
\Pi^{\rm gh}_{\mu_1\mu_2} (x_0, x_{0'})
&= 2 \frac{(x_{00'})_{\mu_1} (x_{00'})_{\mu_2}}{x_{00'}^8}
\, , \\
\Pi^{\rm fs}_{\mu_1\mu_2} (x_0, x_{0'})
&= 
- \frac{4 n_{\rm f} + n_{\rm s}}{N_c}
\left(
g_{\mu_1 \mu_2} - 2 \frac{(x_{00'})_{\mu_1} (x_{00'})_{\mu_2}}{x_{00'}^2}
\right)
\, ,
\end{align}
correspond to respective diagrams on the right hand-side of the equality sign of Fig.\ \ref{SelfEnergyFig}. The first and second structures agree with
known polarization tensors, respectively, see, e.g., Ref.\ \cite{Pascual:1984zb}. Here, when displaying results for `adjoint matter', we were slightly more 
generic and exhibited them from $n_f$ fundamental fermions and $n_s$ real scalars. In $\mathcal{N} = 4$ SYM, $n_{\rm f} = 4 N_c$ and  $n_{\rm f} = 6 N_c$.


\end{document}